\documentclass[11pt,english,nofootinbib]{revtex4}
\usepackage[T1]{fontenc}
\usepackage[latin9]{inputenc}

\usepackage{amsmath}
\usepackage{graphicx}
\usepackage{esint}
\usepackage{epsf}
\usepackage{mathrsfs}
\usepackage{slashed}

\def\Tr{{\rm Tr}}
\def\be{\begin{equation}}
\def\ee{\end{equation}}

\begin{document}

\title{Functional renormalization with fermions and tetrads}

\author{P. Don\`a} 
\affiliation{International School for Advanced Studies, via Bonomea 265, 34136 Trieste, Italy}
\affiliation{INFN, Sezione di Trieste, Italy}
\author{R.~Percacci} 
\affiliation{International School for Advanced Studies, 
via Bonomea 265, 34136 Trieste, Italy} 
\affiliation{INFN, Sezione di Trieste, Italy}

\begin{abstract}
We investigate some aspects of the renormalization group flow of gravity
in the presence of fermions, which have remained somewhat puzzling so far.
The first is the sign of the fermionic contribution to the running of 
Newton's constant, which depends on details of the cutoff.
We argue that only one of the previously used schemes correctly implements the
cutoff on eigenvalues of the Dirac operator,
and it acts in the sense of screening Newton's constant.
We also show that K\"ahler fermions give the same contribution 
to the running of the cosmological and Newton constant as four Dirac spinors.
We then calculate the graviton contributions to the beta functions
by imposing the cutoffs on the irreducible spin components of the tetrad.
In this way we can probe the gauge dependence of the off-shell flow.
The results resemble closely those of the metric formalism,
except for an increased scheme-- and (off shell) gauge--dependence. 
\end{abstract}
\maketitle

\section{Cutoff schemes}

The functional renormalization group equation (FRGE) is a convenient method
to calculate the quantum effective action that is not restricted to renormalizable field theories \cite{wegner,wetterich}.
Instead of computing directly the effective action, one computes its derivative with
respect to an external mass scale $k$ called ``the cutoff''.
This ``beta functional'' is automatically free of UV and IR divergences.
The advantage of this procedure is that it provides a definition of the
beta functions of a theory that does not refer to any UV regularization.
Then, taking some (generally local) action as starting point at an UV scale $\Lambda$,
one can obtain the effective action by integrating the FRGE from $\Lambda$ to zero.
The divergences of the theory reappear when one tries to shift $\Lambda$ to infinity
and can be analysed by integrating the FRGE in the direction of increasing $k$.

For our purposes the most useful implementation of this idea is the Wetterich equation \cite{wetterich}.
It gives the $k$-dependence of a functional called the Effective Average Action (EAA),
which is a modified effective action where the propagation of the low momentum modes is suppressed. 
Here ``low momentum'' generally means ``momenta below a given scale $k$''.
The suppression is described, in flat space, by some function $R_k(q^2)$, 
where $q^2$ is the squared momentum.
To be admissible, the cutoff function $R_k$ must satisfy the following basic requirements:
(i) it must be continuous and monotonically decreasing in $q^2$ and $k$,
(ii) it must go rapidly to zero for $q^2>k^2$, 
(iii) it must tend to a positive (possibly infinite) value for $q^2\to 0$
and (iv) it must tend to zero for $k\to 0$.
It is unavoidable that details of the flow depend on the shape of the cutoff function,
even though the effective action, which is obtained by letting $k\to0$, is not.
This is referred to as ``scheme dependence'' and is akin to the renormalization scheme
dependence in perturbation theory.

When one considers anything more than a single scalar field,
in addition to the shape of the function $R_k(z)$, one encounters further choices.
We will be concerned mainly with gravity, which offers a particularly complex scenario.
The application of the FRGE to gravity has been pioneered in \cite{reuter1},
followed by \cite{dou}. 
Already in these early works the implementation of the cutoff followed different methods.
In \cite{reuter1} the cutoff was imposed on the tracefree and trace parts of the metric fluctuation 
(a decomposition that is purely algebraic), 
whereas in \cite{dou} the cutoff was imposed separately on each irreducible spin component
(a finer decomposition that involves differential conditions).
Following \cite{cpr2}, we will call the former a cutoff of type ``a'' and 
the latter of type ``b''.
When gravity is involved, the kinetic operators governing the propagation of quantum fields
typically have the structure $-\nabla^2+aR+bE$, where $\nabla$ is the Riemannian covariant derivative,
$R$ stands here for some combination of curvature terms (Riemann, Ricci and scalar curvatures, with indices
arranged in various ways) and $E$ a term that may involve other background fields
and possibly also couplings.
Since Fourier analysis is not available in general curved spaces,
in defining the cutoff one has to specify an operator whose eigenvalues
form a basis in field space and in general play the role that plane waves play in flat space.
One takes $R_k$ to be a function of this operator (or rather, its eigenvalues).
In \cite{cpr2} the following terminology has been introduced:
a cutoff is of type I if $R_k$ is a function of $-\nabla^2$, of type II if it is a function
of $-\nabla^2+aR$ and of type III if it is a function of $-\nabla^2+aR+bE$.
It is clear that there are infinitely many more possible cutoff variants and that this classification
is very incomplete, but it covers the most commonly used cutoffs.

The RG flow in the so--called Einstein--Hilbert truncation, where one retains in the action
terms at most linear in curvature, has been studied in 
\cite{souma,lauscher,saueressig,litim,benedetti},
with en extended ghost sector in \cite{eichhorn,grohs,eichhorn2}
and using the Vilkoviski-de Witt formalism in \cite{donkin}.
Higher derivative terms have been analysed in 
\cite{lauscher2,codello1,niedermaier,cpr1,cpr2,ms,bms1}.
The contribution of matter fields has been taken into account in
\cite{dou,perini1,perini2,narain,bms2} using a Type I cutoff
and in \cite{largen} using a type II cutoff.
Conversely, the effect of gravity on fermionic interactions has been studied in \cite{zzvp,rodigast,zv,eichhorngies}.

Physical observables, which are related to the full effective action
at $k=0$, will be independent of the cutoff choice.
Furthermore, some terms in the beta functional, namely those that refer to dimensionless
couplings, can actually be shown to be scheme--independent.
It is also well--understood that if the system has a nontrivial fixed point,
its position is not universal.
In general, the scheme--dependence of the flow should therefore not be a cause of major concern \cite{narain3}.

In the literature on the renormalization group for gravity, there is however one result
where the scheme dependence seems to be particularly nasty:
it is the contribution of fermion fields to the beta function of Newton's constant.
This contribution has been computed for example in section III of \cite{cpr2}.
If there are $n_D$ Dirac fields it is
\begin{eqnarray}
\label{matterergeI}
-n_D\frac{1}{2}\frac{1}{(4\pi)^2}\int\,d^4x\,\sqrt{g}\,
\left(\frac{1}{6}Q_1\left(\frac{\partial_t R_k}{ P_k}\right)
-\frac{1}{4}Q_2\left(\frac{\partial_t R_k}{P_k^2}\right)\right)4\,R
\end{eqnarray}
if one uses a type I cutoff and
\begin{eqnarray}
\label{matterergeII}
-n_D\frac{1}{2}\frac{1}{(4\pi)^2}\int\,d^4x\,\sqrt{g}\,
\left(\frac{1}{6}-\frac{1}{4}\right)\,Q_1\left(\frac{\partial_t R_k}{P_k}\right)4\,R
=\frac{n_D}{96\pi^2}Q_1\left(\frac{\partial_t R_k}{P_k}\right)\int\,d^4x\,\sqrt{g}\,\,R
\end{eqnarray}
if one uses a type II cutoff. 
Here $Q_{n}(W)=\frac{1}{\Gamma(n)}\int_{0}^{\infty}dz\, z^{n-1}W(z)$.
If we use the optimized cutoff \cite{litim}
\be
\label{opt}
R_k(z)=(k^2-z)\theta(k^2-z)\ ,
\ee
we get $Q_1\left(\frac{\partial_t R_k}{P_k}\right)=2k^2$, 
$Q_2\left(\frac{\partial_t R_k}{P_k^2}\right)=k^2$
and therefore the contribution of fermions to the relevant term of
the beta functional is
\begin{eqnarray}
-\frac{n_D}{96\pi^2}\int\,d^4x\,\sqrt{g}\,R&\qquad \mbox{for the type I cutoff}
\\
\frac{n_D}{48\pi^2}\int\,d^4x\,\sqrt{g}\,R&\qquad \mbox{for the type II cutoff}.
\end{eqnarray}
The latter agrees with an earlier independent calculation 
of the renormalization of $G$ in \cite{larsen};
the former differs not just in the value of the coefficient but even in the sign.
\footnote{This was noted a while ago by Calmet et al. while they were working on \cite{reeb}.
R.P. wishes to thank D. Reeb for correspondence on this point.}
To check that this is not just a quirk of the optimized cutoff, consider an exponential cutoff 
\be
\label{exponentialcutoff}
R_k(z)=\frac{z e^{-a(z/k^2)^b}}{1-e^{-a(z/k^2)^b}}\ .
\ee
In order to guarantee that condition (iii) is satisfied, one has to assume $b\geq1$.
For example if $b=1$, $Q_1\left(\frac{\partial_t R_k}{P_k}\right)=\pi^2k^2/3a$ and
$Q_2\left(\frac{\partial_t R_k}{P_k^2}\right)=2k^2/a$
and therefore we encounter the same problem, independently of the parameter $a$.
One can see that the same is true also for $b>1$.

This sign ambiguity is puzzling. 
Since the beta function of $G$ vanishes for $G=0$, the sign of $G$ can never change in the course of the RG flow (we disregard here as pathological the case when the inverse of $G$ passes through zero).
Therefore, if the physical RG trajectory approaches a fixed point in the UV,
the sign of Newton's constant at low energy will be the same as that of Newton's constant at the fixed point.
In a model where gravity is induced by minimally coupled fermions, the sign of Newton's constant 
would depend on the scheme.

One has to be careful in drawing physical conclusions from these calculations:
the relation between the coupling $G$ appearing in $\Gamma_k$ and the physical Newton's constant
that is measured in the lab may not be as simple as it seems.
The functional that obeys an exact RG equation is not a simple functional of a single metric
but rather a functional of a background metric and (the expectation value of) the fluctuation of
the metric. This functional is not invariant under shifts of the background and fluctuation
that keep the sum constant, and therefore there may be several couplings that one could
legitimately call ``Newton's constant'' (for a discussion see \cite{bimetric}). 
The one that we are discussing here is the one that multiplies the Hilbert action 
constructed from the background metric only, and it is not obvious that the
other ones would behave in the same way.
In order to make a completely well--defined statement one should really calculate a physical
observable and in such a calculation all ambiguities should disappear.
It is therefore possible, in principle, that even the sign difference
between the two calculations discussed above may be resolved when one 
considers physical observables.
In this paper we will argue for a different and simpler solution of the issue,
namely that only the type II cutoff gives a result with the physically correct sign.

As an aside, we will also calculate the contribution to the beta functions
of the cosmological constant and Newton constant
due to Euclidean K\"ahler fermions in four dimensions. 
K\"ahler fermions are a way of describing fermion fields in terms of Grassmann algebra elements, 
instead of spinors and has the merit that it does not require the use of tetrad fields.
We find that one K\"ahler fermion gives exactly the same contribution as four Dirac spinors,
and that the same sign issue is present.

When gravity is coupled to fermions an additional question arises regarding the
field carrying the gravitational degrees of freedom.
In \cite{dou,perini1,perini2}, where the contribution of graviton loops was added to that of matter fields,
as well as in \cite{zzvp,rodigast,zv} where scalars and fermions were interacting with gravity,
the carrier of the gravitational degrees of freedom was the metric.
It has been pointed out in \cite{hr} that in the presence of fermions it may be more natural to use the tetrad.
Even if one chooses a Lorentz gauge where the antisymmetric part of the tetrad fluctuation 
is suppressed, the two calculations are not the same, because one has to work off shell and
the Hessian in the tetrad formalism contains some additional terms
proportional to the equations of motion.
The calculations in \cite{hr}, which used a type Ia cutoff in the $\alpha=1$ gauge, 
show that a fixed point is still present, but is less stable than in the metric formulation.
We will compute the gravitational contributions in the tetrad formalism, using type b cutoffs
(both Ib and IIb); this allows us to analyse the off--shell $\alpha$--dependence.
The results will be found to be somewhat closer to those of the metric formalism
than those found in \cite{hr} (which we have independently verified),
but they still present a stronger dependence on the scheme and on the gauge parameter.

\section{K\"ahler fermions}

Before discussing in detail the issue of cutoff scheme with ordinary spinor fermions,
we would like to point out that precisely the same problem arises also in another representation
of fermionic matter. This section is not strictly necessary for the following discussion
so readers that are mainly interested in the solution of the puzzle presented above
may go directly to section III.

As is well--known, in any dimension the Grassmann and the Clifford algebras are isomorphic
as vector spaces. This is the basis for a representation of fermion fields as 
inhomogeneous differential forms \cite{Kah,graf,benn}. Such fields are called Kähler fermions.
In this representation the analogue of the Dirac operator is the first order operator
$d+\delta$, where $d$ is the exterior derivative and $\delta$ is its adjoint.
Note that the definition does not require the use of a tetrad.
We would like to compare the contribution of Kähler fermions
to the gravitational beta functions to the one of ordinary spinor fermions.
In particular we would like to see whether the same sign issue arises
when different cutoff types are used.
Since the details of the calculation are strongly dimension--dependent,
we shall restrict our attention to the case $d=4$.

An inhomogeneous complex differential form $\Phi$ can be expanded as
\be
\Phi=\varphi(x)+\varphi_{\mu}(x)dx^{\mu}+\frac{1}{2!}\varphi_{\mu\nu}(x)dx^{\mu}\wedge dx^{\nu}+\frac{1}{3!}\varphi_{\mu\nu\rho}(x)dx^{\mu}\wedge dx^{\nu}\wedge dx^{\rho}+\frac{1}{3!}\varphi_{\mu\nu\rho\sigma}(x)dx^{\mu}\wedge dx^{\nu}\wedge dx^{\rho}\wedge dx^{\sigma}
\nonumber
\ee
We can map the 3- and 4-form via Hodge duality into a 1- and 0-form respectively.
The field $\Phi$ thus describes a scalar, a pseudoscalar, a vector, a pseudovector and an antisymmetric tensor, 
for a total of 16 complex components.
This is an early sign of the fact that one K\"ahler field is equivalent to four Dirac fields. 

The square of the K\"ahler operator is the Laplacian on forms:
$\Delta=(d+\delta)^2=d\delta+\delta d$.
On forms of degree 0, 1 and 2 it is given explicitly by
\begin{align}
\Delta^{(0)} & =-\nabla^2
\\
\Delta^{(1)\mu}_\nu & =-\nabla^2\delta^\mu_\nu+R_{\phantom{\mu}\nu}^\mu
\\
\Delta^{(2)}_{\alpha\beta}{}^{\gamma\sigma} & =
-\nabla^2\mathbf{1}_{\alpha\beta}{}^{\gamma\sigma}
+R_{\alpha}^{\phantom{\alpha}\gamma}\delta_{\beta}^{\phantom{\alpha}\sigma}
-R_{\beta}^{\phantom{\alpha}\gamma}\delta_{\alpha}^{\phantom{\alpha}\sigma}
-2R_{\alpha\phantom{\gamma}\beta}^{\phantom{\alpha}\gamma\phantom{\beta}\sigma}
\end{align}
In order to read off the beta functions of the cosmological constant and Newton's constant it
enough to consider a spherical (Euclidean de Sitter) background, with curvature tensor
\be
R_{\mu\nu\rho\sigma}=\frac{1}{d\left(d-1\right)}(g_{\mu\rho}g_{\nu\sigma}-g_{\mu\sigma}g_{\nu\rho})R\ ;\qquad
R_{\mu\nu}=\frac{1}{d}g_{\mu\nu}R
\ee
Then the operators defined above reduce to
\begin{align}
\Delta^{(1)\mu}_\nu & =\left(-\nabla^2+\frac{1}{4}R\right)\delta^\mu_\nu
\\
\Delta^{(2)}_{\alpha\beta}{}^{\gamma\sigma} & =
\left(-\nabla^2+\frac{1}{3}R\right)
\mathbf{1}_{\alpha\beta}{}^{\gamma\sigma}
\end{align}

The ERGE for a K\"ahler fermion with a type II cutoff is
\begin{align}
\frac{d\Gamma_{k}}{dt}=& 
-2\,\frac{1}{2}{\rm Tr}_{(0)}\left(\frac{\partial_{t}R_{k}(\Delta^{(0)})}{P_{k}(\Delta^{(0)})}\right)
-2\,\frac{1}{2}{\rm Tr}_{(1)}\left(\frac{\partial_{t}R_{k}(\Delta^{(1)})}{P_{k}(\Delta^{(1)})}\right)
-\frac{1}{2}{\rm Tr}_{(2)}\left(\frac{\partial_{t}R_{k}(\Delta^{(2)})}{P_{k}(\Delta^{(2)})}\right)
\nonumber\\
= & 4\cdot\frac{1}{2}\frac{1}{(4\pi)^{2}}\int\, d^{4}x\,\sqrt{g}\left[-4\, Q_{2}\left(\frac{\partial_{t}R_{k}}{P_{k}}\right)+\frac{1}{3}R\, Q_{1}\left(\frac{\partial_{t}R_{k}}{P_{k}}\right)\right]\ .
\end{align}
For a type I cutoff we find instead:
\begin{align}
\frac{d\Gamma_{k}}{dt}=& -2\cdot\frac{1}{2}{\rm Tr}_{(0)}\left(\frac{\partial_{t}R_{k}(-\nabla^2)}{P_{k}(-\nabla^2)}\right)-2\cdot\frac{1}{2}{\rm Tr}_{(1)}\left(\frac{\partial_{t}R_{k}(-\nabla^2)}{P_{k}(-\nabla^2)+\frac{R}{4}}\right)
-\frac{1}{2}{\rm Tr}_{(2)}\left(\frac{\partial_{t}R_{k}(-\nabla^2)}{P_{k}(-\nabla^2)+\frac{R}{3}}\right)
\nonumber\\
= &\, 4\cdot\frac{1}{2}\frac{1}{(4\pi)^{2}}\int\, d^{4}x\,\sqrt{g}\left[-4\, Q_{2}\left(\frac{\partial_{t}R_{k}}{P_{k}}\right)-R\left(\frac{2}{3}\, Q_{1}\left(\frac{\partial_{t}R_{k}}{P_{k}}\right)-Q_{2}\left(\frac{\partial_{t}R_{k}}{P_{k}^{2}}\right)\right)\right]\ .
\end{align}
Evaluating the $Q$--functionals for the cutoff (\ref{opt}) we get:
\begin{align}
\text{Type II:}\quad&\frac{d\Gamma_{k}}{dt}=4\cdot\frac{1}{2}\frac{1}{(4\pi)^{2}}\int\, d^{4}x\,\sqrt{g}\left[-4\, k^{4}+\frac{2}{3}R\, k^{2}\right]\\
\text{Type I: }\quad&\frac{d\Gamma_{k}}{dt}=4\cdot\frac{1}{2}\frac{1}{(4\pi)^{2}}\int\, d^{4}x\,\sqrt{g}\left[-4\, k^{4}-\frac{1}{3}R\, k^{2}\right]
\end{align}
In both cases the effect of one K\"ahler fermion is seen
to match exactly the result of four spinors ($n_D=4$).
This should not induce one to believe that spinors and K\"ahler fermions are equivalent:
in fact, their contributions to the curvature squared terms are quite different.
Nevertheless, the puzzling sign issue of the $R$--term that afflicts spinor fermions is
present in this case too.

\section{Cutoff choice for fermions}

We now return to ordinary Dirac spinor fields and we reexamine
in more detail their contribution to the gravitational effective action
and beta functions. For the sake of generality we will now work in arbitrary dimension $d$.
The standard way of defining the effective action for a fermion field
that is minimally coupled to gravity
is to exploit the properties of the logarithm and write
\be
\Gamma=-\Tr\log(|\slashed D|)=-\frac{1}{2}\Tr\log(\slashed D^2)
=-\frac{1}{2}\Tr\log\left(-\nabla^2+\frac{R}{4}\right)\ .
\ee
The corresponding EAA can then be defined as
\be
\label{squaredeaa}
\Gamma_k=-\frac{1}{2}\Tr\log\left(-\nabla^2+\frac{R}{4}+R_k\right)\ .
\ee
In the definition of this functional one encounters the same ambiguities that we have 
mentioned earlier for bosonic systems.
In addition to the shape of the cutoff function $R_k$,
one seems to also have the freedom of choosing the argument of this function
to be either $-\nabla^2$ (type I cutoff) or $-\nabla^2+\frac{R}{4}$ (type II cutoff).
The former choice has been made in \cite{dou,perini1,perini2,narain,bms2}, the latter in \cite{largen}.
Taking the $t$--derivative (where $t=\log(k/k_0)$) and defining $P_k(z)=z+R_k(z)$, one has
\begin{eqnarray}
\label{spinorI}
\frac{d\Gamma_k}{dt}=-\frac{1}{2}\Tr\frac{\partial_tR_k(-\nabla^2)}{P_k(-\nabla^2)+\frac{R}{4}}
&\qquad\mbox{for a type I cutoff}
\\
\label{spinorII}
\frac{d\Gamma_k}{dt}=-\frac{1}{2}\Tr\frac{\partial_tR_k(-\nabla^2+\frac{R}{4})}{P_k(-\nabla^2+\frac{R}{4})}
&\qquad\mbox{for a type II cutoff}.
\end{eqnarray}
The first few terms in the curvature expansion of these traces can be evaluated,
for any background, using heat kernel methods.
However, for a spherical background, the spectrum of the Dirac operator is known explicitly
and the same traces can also be computed directly as spectral sums.
Comparison of these calculations indicates that only the type II cutoff
correctly reflects the cutoff on eigenvalues of the Dirac operator.

\subsection{Heat kernel evaluation}

\label{HeatKernelsection}
With a type I cutoff, the trace (\ref{spinorI})
giving contribution of $n_D$ Dirac spinors to the FRGE is
\be
\label{matterergeI}
\frac{d\Gamma_k}{dt}=
-\frac{n_D}{2}\frac{2^{[d/2]}}{(4\pi)^{d/2}}\int\,d^dx\,\sqrt{g} \Biggl[
Q_{\frac{d}{2}}\left(\frac{\partial_t R_k}{P_k}\right)
+\left(\frac{1}{6}Q_{\frac{d}{2}-1}\left(\frac{\partial_t R_k}{ P_k}\right)
-\frac{1}{4}Q_{\frac{d}{2}}\left(\frac{\partial_t R_k}{P_k^2}\right)\right)R 
+\ldots
\Biggr]\ .
\ee
Here $2^{[d/2]}$ (where $[x]$ is the integer part of $x$) is the dimension of the representation. 
The first term proportional to $R$ is proportional to the heat kernel coefficient $b_2(-\nabla^2)$,
and the second comes from the expansion of the denominator in (\ref{spinorI}).
Evaluating the $Q$--functionals with the cutoff (\ref{opt}) we obtain
\be
\label{FRGEhkI}
\frac{d\Gamma_k}{dt}=-\frac{1}{\Gamma\left(\frac{d}{2} +1 \right)}
\frac{2^{[d/2]}}{(4\pi)^{d/2}} n_D\int \,d^d x\,\sqrt{g} 
\left[k^d+\frac{d-3}{12} k^{d-2}R\right]\ .
\ee

Evaluating (\ref{spinorII}) with the same techniques yields
\be
\label{matterergeII}
\frac{d\Gamma_k}{dt}=
-\frac{n_D}{2}\frac{2^{[d/2]}}{(4\pi)^{d/2}}\int\,d^dx\,\sqrt{g} \Biggl[
Q_{\frac{d}{2}}\left(\frac{\partial_t R_k}{P_k}\right)
-\frac{1}{12}R\,Q_{\frac{d}{2}-1}\left(\frac{\partial_t R_k}{ P_k}\right)
+\ldots \Biggr]\ ,
\ee
where the term proportional to $R$ comes entirely from the heat kernel coefficient
$b_2\left(-\nabla^2+\frac{R}{4}\right)$.
Evaluating with the cutoff (\ref{opt}) we obtain
\be
\label{FRGEhkII}
\frac{d\Gamma_k}{dt} =
-\frac{1}{\Gamma\left(\frac{d}{2} +1 \right)} 
\frac{2^{[d/2]}}{(4\pi)^{d/2}}n_D\int \,d^d x\,\sqrt{g} \left[k^{d}
-\frac{d}{24}k^{d-2}R\right]\ .
\ee
In $d=4$ this yields the results quoted in section I.
We see that the sign issue is present in any dimension $d>3$.

\subsection{Spectral sums on $S^d$}

The heat kernel calculation of the preceding subsection can be done in an arbitrary background.
On the other hand, in the case of the spherical background we know explicitly
the spectrum of the Dirac operator: the eigenvalues and multiplicities are
\be
\lambda_{n}^{\pm}=\pm\sqrt{\frac{R}{d(d-1)}}\left(\frac{d}{2}+n\right)\ ,
\qquad 
m_{n}=2^{\left[\frac{d}{2}\right]}\left(\begin{array}{c}
n+d-1\\
n
\end{array}\right)\ ,\qquad n=0,1,\ldots\ .
\ee
With this information one can compute the trace of any function of the Dirac operator
as $\Tr\,f(\slashed D)=\sum_{n=0}^\infty m_n f(\lambda_n)$.
We will now evaluate the r.h.s. of the FRGE by imposing a cutoff on the
eigenvalues of the Dirac operator. 
The EAA can be defined directly in terms of the Dirac operator as
\be
\label{lineareaa}
\Gamma_k=-\mathrm{tr}\log\left(|\slashed D|+R_k^D(|\slashed D|)\right)\ ,
\ee
where the cutoff $R_k^D$ has to be a function of the modulus of the Dirac operator,
since we want to suppress the modes depending on the wavelength of the corresponding
eigenfunctions. This is also needed for reasons of convergence.
Since the operator is first order, the conditions that $R_k^D$ has
to satisfy are similar to (i)-(iv) of section I, except for the
replacement of $q^2$ and $k^2$ by $\lambda_n$ and $k$ respectively.
For the explicit evaluation, we will use the optimized profile
\be
\label{lincutoff}
R_k^D(z)=(k-z)\theta(k-z)\ ,\ \ \ (z>0)\ .
\ee 
Then we have
\be
\label{sums}
\mathrm{Tr}\left[\frac{\partial_{t}R_k^D(|\slashed{D}|)}{P_k^D(|\slashed D|)}\right]
=\sum_n m_n\frac{\partial_{t}R_{k}^D(|\lambda_n|)}{P_{k}^D(|\lambda_n|)}
=\sum_\pm\sum_n m_n\theta(k-|\lambda_n|)\ .
\ee
The sum can be computed using the Euler-Maclaurin formula.
Details are given in Appendix I.
The result is
\be
\frac{d\Gamma_k}{dt}=
-\mathrm{Tr}\left[\frac{\partial_{t}R_k^D}{P_k^D}\right]=
-\frac{1}{\Gamma\left(\frac{d}{2}+1\right)}
\frac{2^{\left[\frac{d}{2}\right]}}{\left(4\pi\right)^{\frac{d}{2}}}
V(d)
\left(k^{d}-\frac{d}{24}k^{d-2}R+O\left(R^{2}\right)\right)\ ,
\ee
where $V(d)$ is the volume of the $d$--sphere.
This agrees exactly with (\ref{FRGEhkII}), which was obtained with the type II cutoff.
(We have checked that the agreement extends also to the next order in the curvature expansion.)

\subsection{Discussion}

Note that computing the r.h.s. of the FRGE with a spectral sum is a much more direct procedure,
since it avoids going through the square root of the square of the Dirac operator,
and also avoids having to use the Laplace transform and the heat kernel.
It is therefore also a more reliable procedure when there are ambiguities.
The agreement of the spectral sum with the type II--heat kernel calculation
is a useful consistency check and suggests that the latter gives the correct result
whereas the type I cutoff does not.

If so, there remain to understand why the type I cutoff should not be admissible in this case.
One can get some hint by thinking of what this cutoff does in terms of eigenvalues of
the Dirac operator.
We begin by noting that (\ref{lineareaa}) can be rewritten as follows:
\footnote{This is a formal relation because the functional $\Gamma_k$ is ill--defined,
but one can write a corresponding relation for $\partial_t\Gamma_k$, with the same result.}
\be
\Gamma_k=-\frac{1}{2}\mathrm{tr}\log\left(|\slashed D|+R_k^D(|\slashed D|)\right)^2
=-\frac{1}{2}\mathrm{tr}\log\left(-\nabla^2+\frac{R}{4}+2|\slashed D|R_k^D(|\slashed D|)
+R_k^D(|\slashed D|)^2\right)
\ .
\ee
One can compare this with (\ref{squaredeaa}).
Note that the cutoff $R_k$ in that formula could be a function of different operators
which, on a sphere, differ by a constant shift.
For the present purposes it is convenient to think of it as a function of $\slashed D^2$.
Calling $\bar R_k$ this function and calling $z=|\slashed D|$, we have
\be
\bar R_k(z^2)=2zR_k^D(z)+R_k^D(z)^2\ .
\ee
We can solve this relation to get
\be
\label{elena}
R_k^D(z)=-z+\sqrt{z^2+\bar R_{k}(z^2)}\ ,
\ee
so for any cutoff imposed at the level of (\ref{squaredeaa}) one can reverse--engineer
an effective cutoff to be imposed at the level of (\ref{lineareaa}) that will give the same result.

In general, this cutoff may fail to satisfy the required conditions, in particular
condition (iv).
For a type II cutoff, $R_k$ in (\ref{squaredeaa}) is a function of $z^2$,
so $\bar R_k(z^2)=R_k(z^2)$. This implies that $\bar R_0(z^2)=0$
and thus also $R_0^D(z)=0$ for all $z>0$.
For a type I cutoff, on the other hand, 
this may not be the case, as we will show in the following examples.

Consider first the optimized cutoffs.
In the type II case one has
$\bar R_k(z^2)=(k^2-z^2)\theta(k^2-z^2)$ and
one finds that in this case the corresponding cutoff $R_k^D(z)$ given by (\ref{elena}) is also
optimized, and precisely of the form (\ref{lincutoff}).
This is a way of understanding why the two calculations give the same result.
In the case of a type I cutoff, we have instead
$\bar R_k(z^2)=(k^2-z^2+R/4)\theta(k^2-z^2+R/4)$, whence one derives
\be
R_k^D(z)=\left(\sqrt{k^2+\frac{R}{4}}-z\right)\theta\left(\sqrt{k^2+\frac{R}{4}}-z\right)\ .
\ee
This does not tend uniformly to zero when $k\to0$.
In the case of an exponential type II cutoff with $\bar R_k=R_k$ given by (\ref{exponentialcutoff}),
we have 
\be
R_k^D(z)=-z+\frac{z}{\sqrt{1-e^{-a(z^2/k^2)^b}}}\ ,
\ee
which has all the desired properties.
On the other hand for an exponential type I cutoff with $\bar R_k$ given by (\ref{exponentialcutoff}),
\be
R_k^D(z)=-z+\sqrt{z^2+\left(z^2-\frac{R}{4}\right)\frac{e^{-a\left(z^2-\frac{R}{4}\right)^b/k^{2b}}}{1-e^{-a\left(z^2-\frac{R}{4}\right)^b/k^{2b}}}}\ .
\ee
For $b$ odd, and in particular for the most natural case $b=1$,
this function does not tend uniformly to zero when $k\to0$ and therefore condition (iv) is not satisfied.
\footnote{Note that in the limit $k\to0$ the function $R_k^D$ is non zero only for $z<\sqrt{R/4}$.
Since the smallest eigenvalue of the Dirac operator is $\sqrt{R/3}$,
it remains true that $\lim_{k\to0}\Gamma_k=\Gamma$.}

These arguments lend support to the view that only the type II cutoff gives
the physically correct result.
Of course not all results obtained from the type I cutoff have to be wrong, for example the
leading term (renormalizing the cosmological constant) and, in $d=4$, the curvature squared terms,
are the same using the two cutoffs.
These however are just the ``universal'' quantities that do not depend on the choice of the cutoff.
We believe that for the generic dimensionful, non--universal quantities, the results obtained
via a type I cutoff should not be trusted.

\section{Tetrad gravity}

We will now compute the graviton contribution to the running
of Newton's constant and cosmological constant, in $d$--dimensions,
when the tetrad is used as a field variable.
This has been discussed recently in \cite{hr} using a type Ia cutoff,
i.e. a cutoff that depends on $-\nabla^2$ that is added to the full gravitational Hessian.
In order to have a manageable, minimal Laplacian--type operator,
this requires that the gravitational gauge--fixing parameter be fixed to $\alpha=1$.
We will use instead cutoffs of type b, meaning that the graviton is first decomposed
into irreducible components of spin 2, 1 and 0, and the cutoff is imposed separately
one each component.
This permits the discussion of general diffeomorphism gauges.
We will use both type Ib and type IIb cutoffs.

\subsection{Hessian and gauge fixing}

The ansatz we make for the effective average action is the standard Einstein--Hilbert truncation
\be
\label{ehaction}
\Gamma_{k}[e,\bar{e}]
=\Gamma^{EH}_{k}[e,\bar{e}]+\Gamma_k^{GF}[e,\bar{e}]
=-\frac{1}{16\pi G_{k}}
\int d^dx\,\det e\,\bigg(R(g(e))-2\Lambda_k\bigg)+\Gamma_k^{GF}[e,\bar{e}]
\ee
where we have indicated the $k$--dependence of the couplings
and $\Gamma_k^{GF}$ is the gauge--fixing term, to be specified below.
In tetrad formulation the metric is represented in terms of vielbeins $e_{\mu}^{a}$ as
$g_{\mu\nu}=e_{\ \mu}^{a}e_{\ \nu}^{b}\eta_{ab}$. 
If we decompose $g_{\mu\nu}\equiv\bar{g}_{\mu\nu}+h_{\mu\nu}$
and $e_{\ \mu}^{a}\equiv\bar{e}_{\ \mu}^{a}+\varepsilon_{\ \mu}^{a}$ we have the relation
\be
h_{\mu\nu}=2\varepsilon_{(\mu\nu)}+\varepsilon_{(\mu}{}^\rho\varepsilon_{\nu)\rho}
\ee
where Latin indices on $\varepsilon$ have been transformed to Greek ones by contraction with
$\bar e$.
Now substituting this formula in the Taylor expansion of the metric in terms
of metric fluctuations we find:
\be
\Gamma^{EH}(e)=\Gamma^{EH}(\bar e)
+\int \frac{\delta\Gamma^{EH}}{\delta g_{\mu\nu}}2\varepsilon_{\mu\nu}
+\zeta\int \frac{\delta\Gamma^{EH}}{\delta g_{\mu\nu}}\varepsilon_{\mu}{}^\rho\varepsilon_{\nu\rho}
+\frac{1}{2}\int\int \frac{\delta\Gamma^{EH}}{\delta g_{\mu\nu}\delta g_{\rho\sigma}}\,4\,\varepsilon_{\mu\nu}\varepsilon_{\rho\sigma}+\ldots
\ee
In the third term we have introduced by hand a factor $0\leq\zeta\leq1$
that interpolates continuously between the pure metric formalism ($\zeta=0$, $h_{\mu\nu}=2\varepsilon_{(\mu\nu)}$)
and the tetrad formalism ($\zeta=1$).
We see that the part of the action quadratic in $\varepsilon$ differs from the one
in the metric formalism by terms proportional to the equations of motion.
Since in the derivation of the beta functions it is essential to work off shell,
we cannot ignore these terms.

The gauge fixing terms for diffeomorphisms and local Lorentz transformations are
\be
\Gamma_k^{GF}[e,\bar{e}]=\frac{1}{2\alpha}\int\mathrm{d}^{\mathrm{d}}x\sqrt{\bar{g}}\bar{g}^{\mu\nu}F_{\mu}F_{\nu}+\frac{1}{2\alpha_{L}}\int\mathrm{d}^{\mathrm{d}}x\sqrt{\bar{g}}\, G^{ab}G_{ab}
\ee
For diffeomorphisms we choose the condition
\be
F_{\mu}=\frac{1}{\sqrt{16\pi G}}\left(\bar{\nabla}^\nu\bar{h}_{\mu\nu}-\frac{\beta}{2}\bar{\nabla}_{\mu}\bar{h}\right)\ ,
\ee
while for the internal $O(d)$ transformation we choose a symmetric vielbein $G^{ab}=\varepsilon^{[ab]}$.
We will choose $\alpha_L=0$ in order to simplify the computation.
This correspond to a sharp implementation of the Lorentz gauge fixing,
where one can simply set $\varepsilon_{[\mu\nu]}=0$
and suppress the corresponding rows and columns in the Hessian.

Next we perform the TT decomposition on the symmetric part of the vielbein fluctuation
\be
\label{TTdecomposition}
\varepsilon_{(\mu\nu)}=h_{\mu\nu}^{TT}+\nabla_{\mu}\xi_{\nu}+\nabla_{\nu}\xi_{\mu}+\nabla_{\mu}\nabla_{\nu}\sigma-\frac{1}{d}g_{\mu\nu}\nabla^{2}\sigma+\frac{1}{d}g_{\mu\nu}h^{2}\ ,
\ee
and the associated field redefinitions $\xi_{\mu}\to\sqrt{-\nabla^{2}-\frac{R}{d}}\,\xi_{\mu}$
and $\sigma\to\sqrt{\left(-\nabla^{2}-\frac{R}{\left(d-1\right)}\right)\left(-\nabla^{2}\right)}\,\sigma$.

With these definitions, and dropping bars from the background quantities for notational
simplicity, the quadratic part of the action (\ref{ehaction}) is
\begin{align}
\label{decTT}
\Gamma_{h^{T}h^{T}}^{(2)}= & \frac{1}{2}\int\sqrt{g}\,\,
h^{T\mu\nu}\left[-\nabla^{2} + \left(\frac{d\left(d-3\right)+4}{d\left(d-1\right)}-\zeta\frac{d-2}{2d}\right)R-(2-\zeta)\Lambda\right]h_{\mu\nu}^{T}\\
\label{decxi}
\Gamma_{\xi\xi}^{(2)}= & \frac{1}{\alpha}\int\sqrt{g}\,\,
\xi^{\nu}\left[-\nabla^{2}+\left(\frac{\alpha\left(d-2\right)-1}{d}-\zeta\alpha\frac{d-2}{2d}\right)R-\alpha\left(2-\zeta\right)\Lambda\right]\xi_{\nu}\\
\label{decsigma}
\Gamma_{\sigma\sigma}^{(2)}= & 
\frac{\left(d-1\right)}{2d}\frac{2\left(d-1\right)-\alpha\left(d-2\right)}{\alpha d}\times
\\
&
\int\sqrt{g}\,\,
\sigma
\left[-\nabla^{2}+\frac{(d-2)(2-\zeta)\alpha-4}{2\left(d-1\right)-\alpha\left(d-2\right)}R-\frac{\alpha d\left(2-\zeta\right)}{2\left(d-1\right)-\alpha\left(d-2\right)}\Lambda\right]\sigma
\nonumber\\
\Gamma_{hh}^{(2)}= & 
\frac{d-2}{4d}\frac{2\left(d^{2}-3d+2\right)\alpha-(d\beta-2)^{2}}{d\alpha(d-2)}\times
\\
&
\int\sqrt{g}\,\, 
h\Big[-\nabla^2
+\frac{\alpha(d-2)\left(d-4+\zeta\right)}{2\left(d^{2}-3d+2\right)\alpha-(d\beta-2)^{2}} R-2\frac{d\alpha\left(d-2+\zeta\right)}{2\left(d^{2}-3d+2\right)\alpha-(d\beta-2)^{2}}\Lambda\Big]h
\nonumber\\
\Gamma_{h\sigma}^{(2)}= & 
\frac{-(d-2)\alpha+d\beta-2}{d\alpha}\,\frac{\left(d-1\right)}{d}
\int\sqrt{g}\,\, 
h\left(-\nabla^{2}-\frac{R}{\left(d-1\right)}\right)\nabla^2\sigma
\end{align}
We notice that for $\beta=\frac{1}{d}\left(\alpha(d-2)+2\right)$ we can get rid of the mixed term.
In the rest of the paper we will work in this ``diagonal'' gauge.
In this case the trace part reduces to
\begin{align*}
\Gamma_{hh}^{(2)} & =-\frac{1}{2}\frac{d-2}{2d}\int\sqrt{g}\, h\left[-\nabla^2 +\frac{d-4+\zeta}{2(d-1)-\alpha(d-2)}R-\frac{2d}{2(d-1)-\alpha(d-2)}\left(1+\frac{\zeta}{d-2}\right)\Lambda\right]h
\end{align*}

After decomposing the diffeomorphism ghost in its transverse and longitudinal parts,
and absorbing $\sqrt{-\nabla^2}$ in the latter, the ghost action is the sum of
\begin{align}
\label{ghostaction}
\Gamma_{\bar{c}_\nu^T c_\mu^T}^{(2)} = \int\sqrt{g}\, \bar{c}_\nu^T \left(\nabla^2 +\frac{R}{d} \right) c_\mu^T \qquad \Gamma_{ \bar{c} c}^{(2)} = \int \,\sqrt{g}\, \bar{c}\left(\nabla^2 +2\frac{R}{d} \right) c
\end{align}
The Lorentz ghosts do not propagate and following standard perturbative procedure
one could neglect them entirely, but we will follow \cite{hr} and introduce a cutoff for them too.
The corresponding contribution to the FRGE is computed, together with other traces, in Appendix II.

\subsection{Beta functions}
The FRGE can now be calculated by introducing a cutoff separately in each spin sector. 
(This is known as a ``type b'' cutoff.)
Using the same heat kernel methods that we have employed in section \ref{HeatKernelsection},
the expression for $\partial_t\Gamma_k$ can be expanded to linear order in $R$.
Comparing the terms of order zero and one in $R$ in the FRGE yields:
\begin{align}
\label{eqconf1}
\partial_{t}\left(\frac{2\Lambda}{16\pi G}\right)= &\, \frac{k^{d}}{16\pi}\left(A_{1}+\eta A_{2}\right)\\
\label{eqconf2}
-\partial_{t}\left(\frac{1}{16\pi G}\right)= &\, \frac{k^{d-2}}{16\pi}\left(B_{1}+\eta B_{2}\right)
\end{align}
where $\eta=-\partial_t G/G$ and $A_i$, $B_i$ are in general polynomials in $\tilde\Lambda=k^{-2}\Lambda$.
From here one can find the beta functions of the dimensionless parameters $\tilde G=k^{d-2}G$ 
and $\tilde\Lambda$:
\begin{align}
\label{betafunctions}
\partial_{t}\tilde G= & (d-2)\tilde G+\frac{B_{1}\tilde G^{2}}{1+\tilde GB_{2}}\\
\partial_{t}\tilde\Lambda= & -2\tilde\Lambda+\tilde G\frac{A_{1}+2B_{1}\tilde\Lambda+\tilde G\left(A_{1}B_{2}-A_{2}B_{1}\right)}{2\left(1+B_{2}\tilde G\right)}
\end{align}
In the following two sections we will give explicit results using specific cutoffs.

For numerical results in $d=4$ we will always use the optimized cutoff (\ref{opt}).
For a discussion of the dependence on the shape of the function $R_k(z)$
we refer to \cite{hr}.
We will instead concentrate on the differences between cutoffs of type I vs. II and type a vs b.
For the type Ia case we refer again to the extensive discussion in \cite{hr},
whose results we have checked independently.
We will report in detail the results for the cases Ib and IIb,
and highlight the differences with the case Ia.

\subsection{Type Ib cutoff}
First we choose as reference operator, in each spin sector, the ``bare'' Laplacian $-\nabla^2$.
The cutoff is a function $R_k(-\nabla^2)$.
This is called a cutoff of type Ib.
The calculation of the coefficients $A_1$, $A_2$, $B_1$ and $B_2$ for arbitrary dimension and cutoff shape 
is described in Appendix A.
Here we just report the result in $d=4$ and for the cutoff profile (\ref{opt}):
\begin{align}
\label{44}
A_1=& \frac{1}{2\pi}\left[
\frac{5}{1-(2-\zeta)\tilde\Lambda}
+\frac{3}{1-\alpha(2-\zeta)\tilde\Lambda}
+\frac{1}{1-\frac{2\alpha(2-\zeta)}{3-\alpha}\tilde\Lambda}
+\frac{1}{1-\frac{2(2+\zeta)}{3-\alpha}\tilde\Lambda}
-8
\right]+A^L(\tilde\mu)
\\
\label{45}
A_2=& \frac{1}{12\pi}
\left[
\frac{5}{1-(2-\zeta)\tilde\Lambda}
+\frac{3}{1-\alpha(2-\zeta)\tilde\Lambda}
+\frac{1}{1-\frac{2\alpha(2-\zeta)}{3-\alpha}\tilde\Lambda}
+\frac{1}{1-\frac{2(2+\zeta)}{3-\alpha}\tilde\Lambda}
\right]
\\
B_1=& \frac{1}{24\pi}\Biggl[
-\frac{20}{1-(2-\zeta)\tilde\Lambda}
-\frac{5 \left(8-3 \zeta\right)}{(1-(2-\zeta ) \tilde{\Lambda})^2}
+\frac{6}{1-\alpha  (2-\zeta ) \tilde{\Lambda}}
+\frac{9(1-\alpha(2-\zeta))}{4(1-\alpha  (2-\zeta ) \tilde{\Lambda} )^2}
\nonumber\\
&
\qquad\ +\frac{4}{1-\frac{2 \alpha  (2-\zeta ) \tilde{\Lambda} }{3-\alpha}} 
+\frac{12+6\alpha(\zeta-2)}{(3-\alpha)\left(1-\frac{2\alpha(2-\zeta)\tilde\Lambda}{3-\alpha}\right)^2}
+\frac{4}{1-\frac{2 (\zeta +2) \tilde{\Lambda} }{3- \alpha}}
-\frac{6\zeta}{(3-\alpha) \left(1-\frac{2 (\zeta +2) \tilde{\Lambda} }{3- \alpha }\right)^2}
\nonumber\\
&
\qquad\ -50\Biggr]+B^L(\tilde\mu)
\end{align}
\begin{align}
B_2=&\frac{1}{144\pi}\left[-\frac{30}{1-(2-\zeta)\tilde{\Lambda}}
-\frac{5 \left(8-3\zeta\right)}{(1-(2-\zeta ) \tilde{\Lambda} )^2}
+\frac{9}{1-\alpha  (2-\zeta ) \tilde{\Lambda}}
+\frac{9(1-\alpha(2-\zeta)}{(1-\alpha  (2-\zeta ) \tilde{\Lambda} )^2}\right.
\\
&\left.
\qquad \ \ +\frac{6}{1-\frac{2 \alpha(2-\zeta ) \tilde{\Lambda} }{3- \alpha }}
+\frac{12+6\alpha(\zeta-2)}{(3-\alpha)\left(1-\frac{2\alpha(2-\zeta)\tilde{\Lambda}}{3-\alpha}\right)^2}
+\frac{6}{1-\frac{2(\zeta +2)\tilde\Lambda}{3-\alpha}}
-\frac{6\zeta }{(3-\alpha )\left(1-\frac{2 (\zeta +2) \tilde{\Lambda} }{3- \alpha }\right)^2}\right]\ .
\nonumber
\end{align}

The result is still quite general: it depends 
on the parameter $\zeta$, which allows us to interpolate continuously between the
purely metric formulation ($\zeta=0$) and the purely tetrad formulation ($\zeta=1$),
on the arbitrary gauge parameter $\alpha$,
which allows us to test the gauge dependence of the results,
and on the parameter $\mu$ that allows us to weigh differently the contribution
of the Lorentz ghosts.

Let us now describe the main features of these flows.
Both in the metric and in the tetrad formulations, a UV--attractive fixed point
is found for all values of $\mu$ and for $\alpha$ not too large.
Its location and the corresponding critical exponents $\vartheta$
(which are defined as minus the eigenvalues of the stability matrix)
are given in table I in the metric ($\zeta=0$) or tetrad ($\zeta=1$) formalism,
in the gauges $\alpha=0$, $\alpha=1$ and with two different values of the dimensionless Lorentz
ghost parameter $\tilde\mu=\mu/k$: $\tilde\mu=\infty$, and $\tilde\mu=1.2$.
The former corresponds to neglecting the Lorentz ghosts entirely and the latter
is chosen to ease comparison with \cite{hr}, who found that this value gives
results that are closest to the metric formalism in the gauge and scheme they use.
Note that the case $\zeta=0$, $\alpha=1$ correspond to the purely metric flow with type Ib
cutoff, which had already been discussed previously in the literature.
Indeed the second row in table I agrees with the third row in table II in \cite{cpr2}.

Whereas with a type Ia cutoff the fixed point becomes UV--repulsive,
and a limit cycle develops, for $\tilde\mu$ sufficiently large,
with the type Ib cutoff it remains UV--attractive for arbitrarily large $\tilde\mu$.
This is a nice feature of this scheme, because it means that the fixed point can also be found
if one adopts the perturbative prescription of neglecting the Lorentz ghosts entirely.
However, the results in the tetrad formalism match most closely those of the
metric formalism when $\tilde\mu$ is chosen to be a bit larger than one.
As with the type Ia cutoff, for $\tilde\mu$ smaller than a critical value $\tilde\mu_c$,
the critical exponents become real. We find $\tilde\mu_c\approx 0.705$
for $\alpha=0$ and $\tilde\mu_c\approx 0.715$ for $\alpha=1$.

The dependence of the universal quantities on the gauge parameters is illustrated in Fig. 1.
The slow decrease of the real part of the critical exponent for $0<\alpha<2$ 
is in agreement with earlier calculations in the metric formalism (see e.g. fig. 9 in \cite{lauscher}).
The results of different schemes seem to converge for $\alpha\to 0$ which, we recall,
is believed to give the physically most reliable picture.
On the other hand when $\alpha$ is greater than some value of order 2
the fixed point becomes repulsive, reproducing the behavior that
had been observed in \cite{hr} for large $\tilde\mu$.
It is tempting to conjecture that also in the cutoff scheme Ia used in \cite{hr}
the fixed point would have the usual properties, even for large $\tilde\mu$,
if one could choose $\alpha$ closer to zero.
The strong $\tilde\mu$--dependence that had been observed there may be due to a
particularly strong $\alpha$--dependence.
Altogether it appears that with a type Ib cutoff, 
the tetrad formalism leads to results that are qualitatively similar 
to those of the metric formalism, and that the correspondence is best when 
$0<\alpha<1$ and the
Lorentz ghosts are turned on, with a parameter $\tilde\mu$ that is a little larger than one. 

\begin{table}
\begin{center}
\begin{tabular}{l|l|l|l|l}
{\rm Scheme}    &   $\tilde\Lambda_*$   &$\tilde G_*$ &$\tilde\Lambda_*\tilde G_*$& $\vartheta$  \\
\hline
Ib, $\zeta=0$, $\alpha=0$                & 0.1569   & 0.9028  & 0.1416 & $2.147\pm2.620 i$ \\
Ib, $\zeta=0$, $\alpha=1$                & 0.1715   & 0.7012  & 0.1203  & $1.689\pm2.486 i$ \\
Ib, $\zeta=1$, $\tilde\mu=\infty$, $\alpha=0$  & 0.2288  & 1.363 &  0.3119 & $2.086\pm2.042 i$   \\
Ib, $\zeta=1$, $\tilde\mu=\infty$, $\alpha=1$  & 0.2478   & 0.9472  &  0.2347 & $0.595\pm3.753 i$ \\
Ib, $\zeta=1$, $\tilde\mu=1.2$, $\alpha=0$   &  0.0691  &  1.518 & 0.1050 & $2.237\pm1.248 i$  \\
Ib, $\zeta=1$, $\tilde\mu=1.2$, $\alpha=1$  & 0.0798    & 1.3196 & 0.1053 & $1.892\pm1.093 i$
\end{tabular}
\end{center}
\caption{The non trivial fixed point in the type Ib cutoff in metric ($\zeta=0$) 
and tetrad ($\zeta=1$) formalism, in the gauges $\alpha=0$ and $\alpha=1$
and with different weights of the Lorentz ghosts.
Recall Re($\vartheta$)>0 implies that the fixed point is UV attractive.}
\label{table2}
\end{table}
\begin{figure}
\begin{minipage}[c]{0.48\linewidth}
\hspace{-1.5cm}
\includegraphics[width=8.5cm]{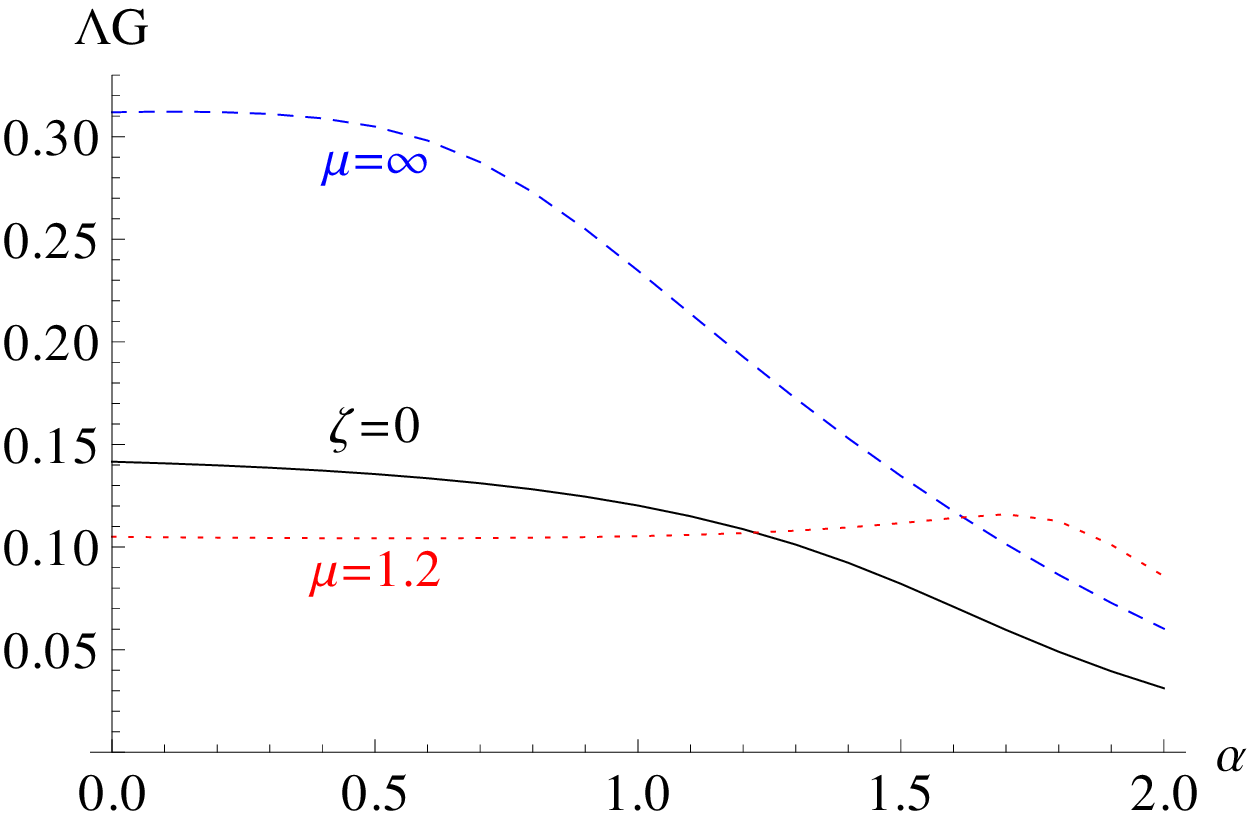} 
\end{minipage}
\hspace{0.02\linewidth}
\begin{minipage}[c]{0.48\linewidth}
\hspace{-1cm}
\includegraphics[width=8.5cm]{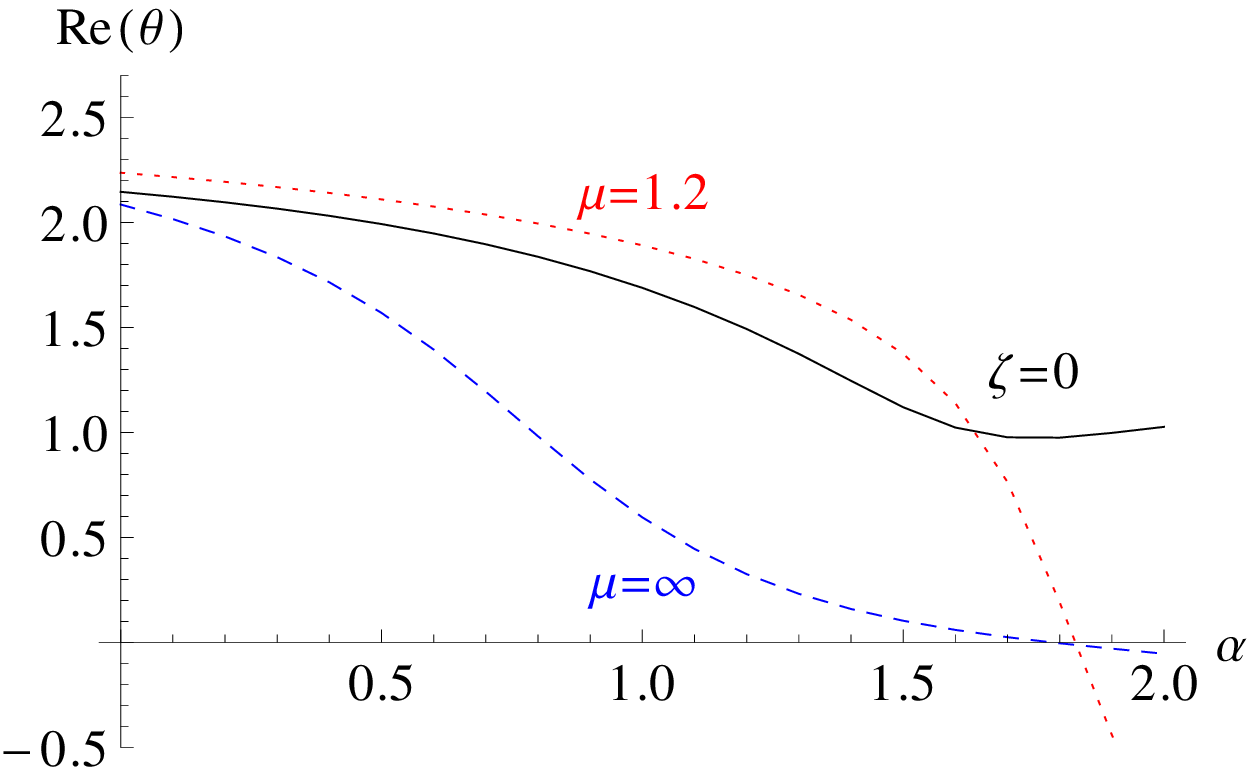} 
\end{minipage}
\caption{Plot of universal quantities as functions of the gauge parameter $\alpha$ for type Ib cutoff. 
In the left panel the product $\Lambda_*G_*$, in the right panel the real part of the critical exponent.}
\end{figure}

\subsection{Type IIb cutoff}

We call type IIb a cutoff imposed separately on each spin--component of the graviton
and taking as reference operator the Laplace--type operator that appears in the
corresponding Hessian, including the curvature terms, but not the term proportional
to the cosmological constant. The rationale for excluding the cosmological constant
term is that the cosmological constant is a running coupling and if one included
it in the reference operator, it would not remain fixed in the course of the flow.
Here we choose a reference operator that remains fixed along the flow.
\footnote{Cutoffs that depend on the full Hessian, including the terms proportional to the cosmological constant, were called of type III in \cite{cpr2}, where they have been applied to the Einstein--Hilbert truncation.}
As we have already seen in the case of the fermions, the type II cutoffs tend to give
somewhat simpler final formulae than the corresponding type I cutoffs, because
to leading order one always finds traces of the function $\partial_t R_k/P_k$
and it is not necessary to expand the denominators.

The coefficients $A_1$, $A_2$ are the same as with a type Ib cutoff and are given in (\ref{44},\ref{45}).
The coefficients $B_1$ and $B_2$ for arbitrary dimension and cutoff shape are given in Appendix III.
In $d=4$ and for the cutoff profile (\ref{opt}), they become
\begin{align}
\label{coeffIIb1}
B_1= & \frac{1}{12\pi}
\left[
-\frac{10(10-3\zeta)}{1-(2-\zeta)\tilde\Lambda}
+\frac{6(4-3\alpha(2-\zeta))}{1-\alpha(2-\zeta)\tilde\Lambda}
+\frac{2-\frac{6\zeta}{3-\alpha}}{1-2\frac{2+\zeta}{3-\alpha}\tilde\Lambda} 
+\frac{14-6\frac{4-\alpha\zeta}{3-\alpha}}{1-2\alpha\frac{2-\zeta}{3-\alpha}\tilde\Lambda}
-40\right]+B^L
\\
\label{coeffIIb2}
B_2=&\frac{1}{48\pi}
\left[
-\frac{5(10-3\zeta)}{1-(2-\zeta)\tilde\Lambda}
-\frac{3(4-3\alpha(2-\zeta)}{1-\alpha(2-\zeta)\tilde\Lambda}
-\frac{2-\frac{6\zeta}{3-\alpha})}{1-2\frac{2+\zeta}{3-\alpha}\tilde\Lambda}
-\frac{14-6\frac{4-\alpha\zeta}{3-\alpha}}{1-2\alpha\frac{2-\zeta}{3-\alpha}\tilde\Lambda}
\right]
\end{align}

Table II gives the UV--attractive fixed point and the corresponding critical exponents
in the metric ($\zeta=0$) or tetrad ($\zeta=1$) formalism,
in the gauges $\alpha=0,1$, and with two different values of the Lorentz
ghost parameter, $\tilde\mu=\infty$, and $\tilde\mu=1.2$.

\begin{table}
\begin{center}
\begin{tabular}{l|l|l|l|l}
{\rm Scheme}    &   $\tilde\Lambda_*$   &$\tilde G_*$ &$\tilde\Lambda_*\tilde G_*$& $\vartheta$  \\
\hline
IIb, $\zeta=0$, $\alpha=0$                & 0.1052   & 0.7216  & 0.0759 & $2.562\pm1.566 i$ \\
IIb, $\zeta=0$, $\alpha=1$                & 0.0924   & 0.5557  & 0.0513  & $2.424\pm1.270 i$ \\
IIb, $\zeta=1$, $\tilde\mu=\infty$, $\alpha=0$  & 0.1406  & 1.0176 &  0.1431 & $2.595\pm1.131 i$   \\
IIb, $\zeta=1$, $\tilde\mu=\infty$, $\alpha=1$  & 0.1369   & 0.8427  &  0.1154 & $2.300\pm0.991 i$ \\
IIb, $\zeta=1$, $\tilde\mu=1.2$, $\alpha=0$   &  0.0394   &  1.0008  & 0.0398  & $2.640\pm0.730 i$  \\
IIb, $\zeta=1$, $\tilde\mu=1.2$, $\alpha=1$  &   0.0361   & 0.8299   & 0.0300  & $2.547\pm0.634 i$
\end{tabular}
\end{center}
\caption{The non trivial fixed point in the type IIb cutoff in metric ($\zeta=0$) 
and tetrad ($\zeta=1$) formalism, in the gauges $\alpha=0$ and $\alpha=1$
and with different weights of the Lorentz ghosts.}
\label{table2}
\end{table}
\begin{figure}
\begin{minipage}[c]{0.48\linewidth}
\hspace{-1.5cm}
\includegraphics[width=8.5cm]{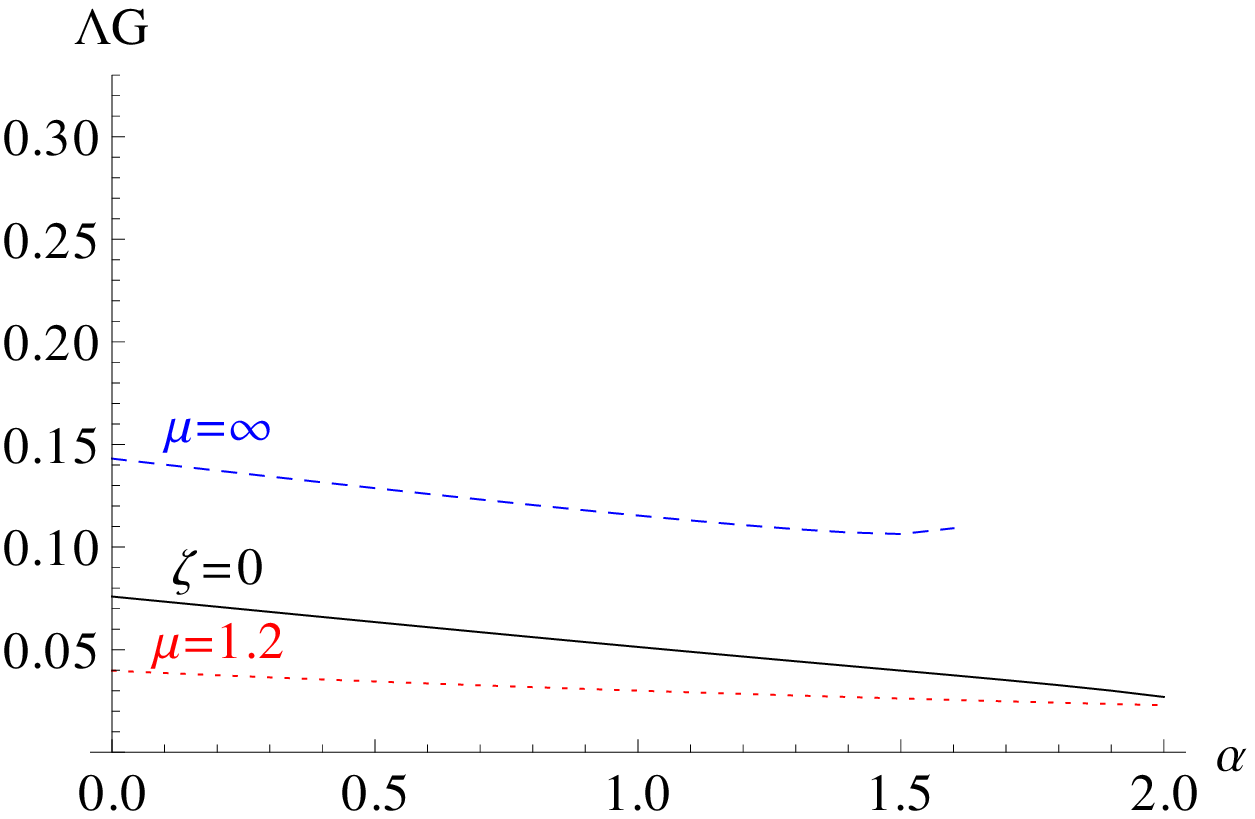} 
\end{minipage}
\hspace{0.02\linewidth}
\begin{minipage}[c]{0.48\linewidth}
\hspace{-1cm}
\includegraphics[width=8.5cm]{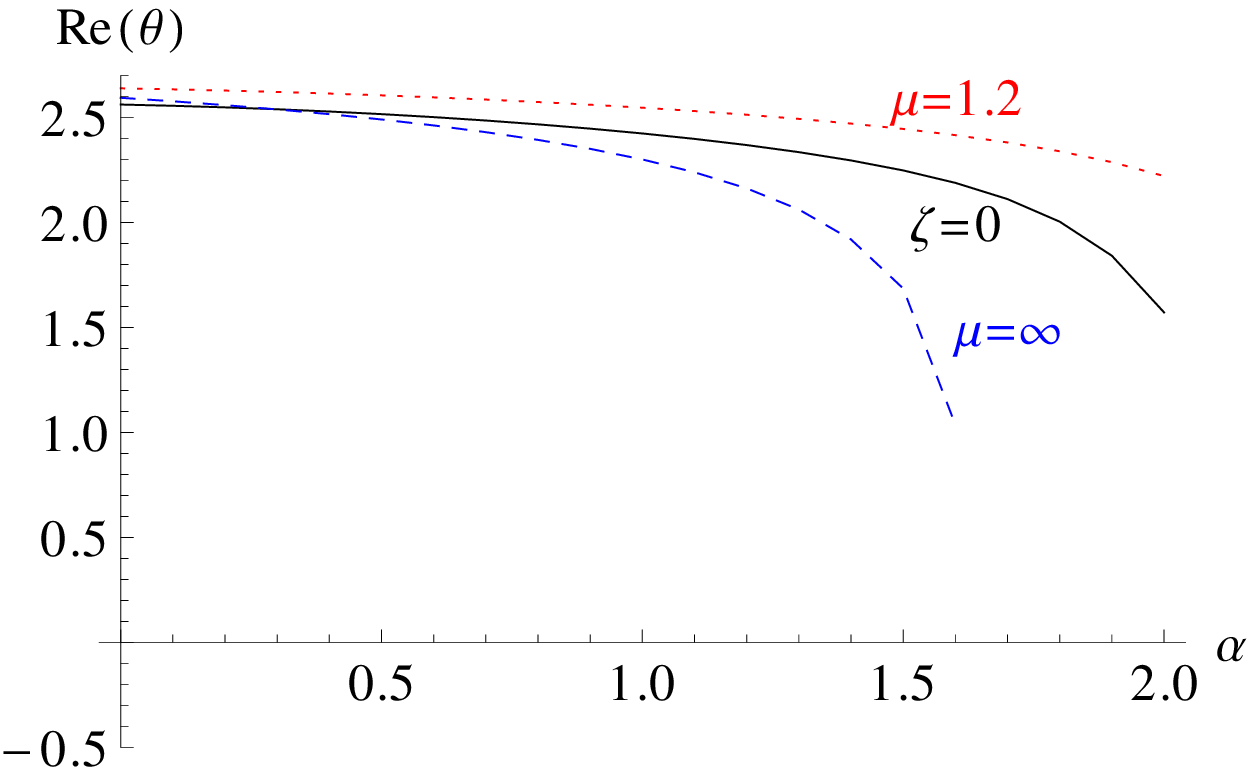} 
\end{minipage}
\caption{Plot of universal quantities as functions of the gauge parameter $\alpha$ for type IIb cutoff. 
In the left panel the product $\Lambda_*G_*$, in the right panel the real part of the critical exponent.}
\end{figure}

The results are qualitatively similar to the ones obtained with the
type Ib cutoff. This is in line with all the results obtained previously
in the Einstein--Hilbert truncation.
The non trivial FP exists and has complex critical exponents for all values of $\tilde\mu$
greater than a critical value $\tilde\mu_c$, which is approximately equal
to 0.766 for $\alpha=0$ and 0.748 for $\alpha=1$.
For small $\tilde\mu$ the FP moves towards negative values of $\tilde\Lambda$.
For large $\tilde\mu$ the fixed point remains UV attractive,
in contrast to the result found in \cite{hr} with the type Ia cutoff scheme
(which we have independently verified).
In particular we find that the FP has properties close to the standard ones
of the metric formulation also when the Lorentz ghosts are neglected.
Figure 2 gives the gauge--dependence of the universal quantities $\Lambda G$ and $\vartheta$.
Note that the real part of the scaling exponent $\vartheta$ is particularly stable
in this scheme, for $0<\alpha<1$.

\subsection{Type IIa cutoff}

For completeness we mention here also the results for the cutoff of type IIa,
which had been discussed first in section IVC of \cite{cpr2}.
In this scheme only the gauge $\alpha=1$ is easily computable.
In this gauge it is enough to split the metric fluctuation into its trace
and tracefree parts to write the Hessian of the Einstein--Hilbert action
as two minimal Laplace--type operators.
The cutoff is then defined as a function of these operators, including the
curvature terms but excluding the cosmological constant term.
This prescription leads to particularly simple expressions.

The coefficients $A_1$, $A_2$ are the same as with a other cutoff types considered here and are given in (\ref{44},\ref{45}).
The coefficients $B_1$ and $B_2$ in $d=4$ and for the cutoff profile (\ref{opt}) are simply
\begin{align}
B_1= &\,\frac{1}{12\pi}
\left[
\frac{2-3\zeta}{1-(2+\zeta)\tilde\Lambda}
-\frac{27(2-\zeta)}{1-(2-\zeta)\tilde\Lambda}
-40\right]+B^L
\\
B_2=&\,\frac{1}{48\pi}
\left[
\frac{2-3\zeta}{1-(2+\zeta)\tilde\Lambda}
-\frac{27(2-\zeta)}{1-(2-\zeta)\tilde\Lambda}
\right]
\end{align}
These expressions coincide with (\ref{coeffIIb1},\ref{coeffIIb2}) when one puts $\alpha=1$ there.
As a consequence, all properties of the flow are the same and we will not discuss this case further.

It is nevertheless interesting to understand the reason for this coincidence, which is not
restricted to $d=4$ and is also independent of the shape of the function $R_k$.
\footnote{The agreement between cutoffs IIa and IIb for $\alpha=1$
had been noticed before in \cite{rahmedethesis}.}
For the sake of simplicity we shall discuss here only the case $\zeta=0$, but the result is general.
Since in all cases the trace field $h$ is treated separately,
and its contribution is the same for type a and b cutoffs,
it is enough to consider the tracefree part of the graviton, namely the components 
$h_{\mu\nu}^{TT}$, $\xi_\mu$ and $\sigma$.
In the gauge $\alpha=1$, the Hessian in the tracefree subsector is a minimal 
second order operator of the form
\be
-\nabla^2+C_TR-2\Lambda\
\ee
where $C_T=\frac{d(d-3)+4}{d(d-1)}$.
When one uses a cutoff of type IIa (no further decomposition)
the contribution of this sector to the r.h.s. of the FRGE is
\be
\label{caroline}
\sum_n\frac{\partial_t R(\lambda_n)-\eta R_k(\lambda_n)}{P_k(\lambda_n)-2\Lambda}
\ee
where $\lambda_n$ are the eigenvalues of the operator $\mathcal{O}=-\nabla^2+C_T R$.
One can divide these eigenvalues into three classes, depending on the spin
of the corresponding eigenfunction.
Upon using the TT decomposition (\ref{TTdecomposition})
one finds that the eigenvalues of $\mathcal{O}$ on fields
of type $h_{\mu\nu}^{TT}$, $\nabla_\mu\xi_\nu-\nabla_\nu\xi_\mu$
and $\nabla_\mu\nabla_\nu\sigma-\frac{1}{d}\nabla^2\sigma$ are equal to the
eigenvalues of the operators in square brackets in (\ref{decTT},\ref{decxi},\ref{decsigma}),
stripped of the $\Lambda$ terms.
We denote these operators $\mathcal{O}^{TT}=-\nabla^2+C_T R$, 
$\mathcal{O}^\xi=-\nabla^2+C_\xi R$ and 
$\mathcal{O}^\sigma=-\nabla^2+C_\sigma R$
and the corresponding eigenvalues $\lambda_n^{TT}$, $\lambda_n^\xi$ and $\lambda_n^\sigma$.
So, the trace (\ref{caroline}) is equal to
\be
\sum_n\frac{\partial_t R(\lambda^{TT}_n)-\eta R_k(\lambda^{TT}_n)}{P_k(\lambda^{TT}_n)-2\Lambda}
+\sum_n\frac{\partial_t R(\lambda^\xi_n)-\eta R_k(\lambda^\xi_n)}{P_k(\lambda^\xi_n)-2\Lambda}
+\sum_n\frac{\partial_t R(\lambda^\sigma_n)-\eta R_k(\lambda^\sigma_n)}{P_k(\lambda^\sigma_n)-2\Lambda}
\ee
Since for $\alpha=1$ the coefficients of $\Lambda$ in (\ref{decTT},\ref{decxi},\ref{decsigma})
are all the same and equal to $-2$,
this is recognized as the contribution of the fields $h_{\mu\nu}^{TT}$, $\xi_\mu$ and $\sigma$ to
the r.h.s. of the FRGE when one uses a cutoff of type IIb.
By a similar reasoning one also concludes that the ghost contribution is the same
in the IIa and IIb schemes.

Things do not work in the same way for type I cutoffs, {\it i.e.} when the cutoff
is a function of $-\nabla^2$.
For a type Ia cutoff the contribution of the tracefree sector to the r.h.s. of the FRGE is 
\be
\label{caroline}
\sum_n\frac{\partial_t R(\lambda_n)-\eta R_k(\lambda_n)}{P_k(\lambda_n)+C_T R-2\Lambda}\ ,
\ee
where $\lambda_n$ now denote the eigenvalues of $-\nabla^2$.
This can be expanded as
\be
\sum_n^{TT}\frac{\partial_t R(\lambda_n)-\eta R_k(\lambda_n)}{P_k(\lambda_n)+C_T R-2\Lambda}
+\sum_n^\xi\frac{\partial_t R(\lambda_n)-\eta R_k(\lambda_n)}{P_k(\lambda_n)+C_T R-2\Lambda}
+\sum_n^\sigma\frac{\partial_t R(\lambda_n)-\eta R_k(\lambda_n)}{P_k(\lambda_n)+C_T R-2\Lambda}\ ,
\ee
where $\sum^{TT}$, $\sum^\xi$ and $\sum^\sigma$ denote the sum over eigenvalues of $-\nabla^2$
on $h_{\mu\nu}^{TT}$, $\nabla_\mu\xi_\nu-\nabla_\nu\xi_\mu$ and 
$\nabla_\mu\nabla_\nu\sigma-\frac{1}{d}\nabla^2\sigma$
respectively.
On the other hand for a type Ib cutoff the same contribution is
\be
\sum_n^{TT}\frac{\partial_t R(\lambda_n)-\eta R_k(\lambda_n)}{P_k(\lambda_n)+C_T R-2\Lambda}
+\sum_n^\xi\frac{\partial_t R(\lambda_n)-\eta R_k(\lambda_n)}{P_k(\lambda_n)+C_\xi R-2\Lambda}
+\sum_n\frac{\partial_t R(\lambda_n)-\eta R_k(\lambda_n)}{P_k(\lambda_n)+C_\sigma R-2\Lambda}
\ee
One clearly sees that the two traces are different.

\section{Discussion}

The implementation of the FRGE in the presence of fermions and gravity presents some subtleties
that had not been fully appreciated until recently.
The sign ambiguity of the fermionic contribution to the running of Newton's constant
had been known for a while, but it was regarded as just another aspect of the
scheme dependence that is intrinsic to applications of the FRGE, albeit a particularly worrying one.
Although a completely satisfactory understanding can only come from a treatment of physical observables,
we have argued here that the correct treatment of fermion fields, when the Dirac operator is squared, 
is to use a cutoff that depends on $-\nabla^2+\frac{R}{4}$ (type II cutoff).
There also follows from our discussion that using a cutoff that depends on $-\nabla^2$ (type I cutoff)
may give physically incorrect results.
Unfortunately, several earlier studies (in particular \cite{perini1,perini2
})
have used this scheme, so some of those results may have to be revised. 
We plan to return to this point in a future publication.

Another issue is the use of tetrad vs. metric degrees of freedom.
We have extended the analysis of tetrad gravity initiated in \cite{hr}
by using a different cutoff (type Ib and IIb vs. type Ia) which allowed us to keep
the diffeomorphism gauge parameter $\alpha$ arbitrary.
We have found that the results for the running couplings using the tetrad formalism
are qualitatively similar to those of the metric formalism, with some quirks.
The following points should be noted.
\\
{\bf (i)} For $\zeta=0$ one recovers the standard metric formalism.
In the case of the type Ib cutoff the results agree with the ones obtained earlier in the literature
\cite{dou,lauscher,cpr2}. 
The type IIb cutoff with generic $\alpha$ had never been used before and the results obtained here are new.
We have shown that for $\alpha=1$ they coincide with the ones found in \cite{cpr2} 
for the type IIa cutoff.
For other values of $\alpha$ they differ only marginally from those obtained with other cutoff types 
and confirm the stability of the fixed point in the metric formalism.
Type II cutoffs have the attractive feature that they lead to somewhat more compact expressions
for the beta functions.
\\
{\bf ii)} The case $\zeta=1$ corresponds to the tetrad formalism.
In this case a new ambiguity appears in the definition of the ghost sector: it can be parameterized by
a mass $\mu$ that appears in the mixing between diffeomorphism and Lorentz ghosts,
or by the corresponding dimensionless parameter $\tilde\mu=\mu/k$.
This parameter is a priori arbitrary, but in order not to introduce additional mass scales
into the problem it is natural to assume that it is of order one.
On the other hand we recall that in perturbation theory and in the chosen gauge the Lorentz ghosts
are neglected, since they do not propagate.
This corresponds to taking $\tilde\mu=\infty$.
\\
{\bf iii)} If one uses a type Ia cutoff and completely neglects the Lorentz ghosts,
there is no attractive FP for positive $G$ \cite{hr}.
Instead, one finds a UV--repulsive fixed point surrounded by a UV--attractive limit cycle\footnote{This is different from the behaviour of the limit cycle discussed in \cite{satz}.}. 
This is not the case when one imposes the cutoff separately on each spin component,
as we did here. We find that both with type Ib and IIb cutoffs an attractive
FP with complex critical exponents is present also when Lorentz ghosts are neglected,
both for $\alpha=1$ and $\alpha=0$. This is reassuring because in the metric formalism the
fixed point can be found even using the perturbative one loop beta functions.
\\
{\bf iv)} If the contribution of Lorentz ghosts is added, as advocated in \cite{hr},
its effect is weighted by the parameter $\tilde\mu$:
it is strong for small $\tilde\mu$ and weak for large $\tilde\mu$.
Since the ghosts are fermions, the fixed point is shifted towards negative $\Lambda$
for decreasing $\tilde\mu$. In addition, they have a systematic effect on the critical exponents:
the modulus of the imaginary part decreases with decreasing $\tilde\mu$ and there is
a critical value $\tilde\mu_c$ under which the critical exponents become real. 
In the gauge $\alpha=1$, $\tilde\mu_c=0.715$ for a type Ib cutoff, $\tilde\mu_c=0.748$ for a type IIb cutoff,
and $\tilde\mu_c\approx0.8$ for a type Ia cutoff.
Similar behaviour is observed also for $\alpha=0$.
On the other hand, the real part of the critical exponent decreases with increasing $\tilde\mu$.
With the type Ia cutoff this effect is most dramatic: the real part becomes negative for $\tilde\mu\approx 1.4$,
and this marks the appearance of the limit cycle.
With the type II cutoffs discussed here the effect is much weaker and the fixed point becomes
only slightly less attractive even in the limit $\tilde\mu\to\infty$,
both for $\alpha=1$ and $\alpha=0$. 
\\
{\bf v)} 
The closest match between the tetrad and metric results is typically obtained
if one chooses some value of $\tilde\mu$ that is not too far from one.
This is always the case for the product $\Lambda G$,
and in most cases also for the critical exponent.
For the Ia cutoff this value was found to be approximately 1.2.
For the Ib and IIb cutoffs it is somewhat larger, depending on the quantities one is comparing.
An exception occurs for the critical exponent in the case of a cutoff of type IIb in the gauge $\alpha=0$,
for which the best match occurs for $\tilde\mu\to\infty$.
\\
{\bf vi)} Using type b cutoffs ({\it i.e.} decomposing the fields into irreducible components)
has the advantage that one can keep the diffeomorphism gauge parameter $\alpha$ arbitrary. 
The gauge dependence of the critical exponents is similar to what had been
observed previously in the metric formalism, as long as $\alpha$ is not too much greater than one.
In particular the real part of the critical exponent tends to decrease as $\alpha$ increases,
for small values of $\alpha$.
In the limit $\alpha\to0$ the $\tilde\mu$--dependence becomes very weak and
the critical exponents nicely converges towards a common value.
On the other hand for $\alpha$ somewhat larger than one the fixed point either becomes
complex or repulsive. This is the behaviour that had been observed in \cite{hr}
with the type Ia cutoff. This suggests that if we were able to compute the
beta functions for this cutoff type with $\alpha\not=1$ we would find that also with large $\tilde\mu$
the fixed point is present and has the standard properties for $\alpha$ sufficiently close to zero.
\\
{\bf vii)} Altogether the results are very similar to those found in the metric formalism,
except for the dependence on the new parameter $\tilde\mu$, which is particularly
strong for the type Ia cutoff.
As argued in \cite{hr}, one can probably attribute the increased sensitivity
of the results to the fact that in the tetrad formalism one has to deal with 
more unphysical degrees of freedom.
The type Ia cutoff seems to be particularly sensitive to the off--shell, 
unphysical features of the flow.
Type b cutoffs, where each spin component is treated separately, are less sensitive.
\\
{\bf viii)} All of the preceding discussion is in the context of a ``single metric truncation'',
{\it i.e.} one assumes that the VEV of the fluctuation field is zero.
As discussed in \cite{bimetric}, the application of the FRGE to gravity requires
that the effective average action be considered in general a function of two variables.
We plan to consider these more general truncations in a future paper.

In conclusion, let us comment on the use of tetrad vs. metric variables.
Since fermions exists in nature, it may seem in principle inevitable that gravity
has to be described by tetrads. This would complicate the theory significantly.
Every diffeomorphism--invariant functional of the metric can be viewed as a 
diffeomorphism and local Lorentz--invariant functional of the tetrad,
but in the bimetric formalism there are many functionals of two tetrads
that cannot be viewed as functionals of two metrics.
Therefore, as already noted in \cite{hr}, the tetrad theory space is
much bigger than the metric theory space.

The necessity of using tetrads should, however, not be taken as a foregone conclusion.
First of all, it is possible that the fermions occurring in nature are K\"ahler fermions.
This would completely remove the argument for the use of tetrads, even in principle.
Whether this is the case or not is a difficult issue that should be answered experimentally.
For the time being one might just consider the use of K\"ahler fermions as a computational trick.
Second, even if we stick to spinor matter, by squaring the Dirac operator 
and using a type II cutoff one can calculate
many quantum effects due to fermions without ever having to use tetrad fields.
The additional complications due to the Lorentz gauge fixing 
and the increased sensitivity to gauge and scheme choice
advise against the use of the tetrad formalism, as a matter of practical convenience.

\bigskip
\centerline{\bf Acknowledgements}
R.P. thanks U. Harst, M. Reuter and F. Saueressig for discussions and hospitality at the University of Mainz
during the preparation of this paper.

\section{Appendix I: Dirac spectral sums}

To compute the sum (\ref{sums}) we can use the Euler-~Maclaurin formula
\be
\sum_{i=0}^{n}F(i)=\int_{0}^{n}F(x)\, dx-B_{1}\cdot(F(n)+F(0))+\sum_{k=1}^{p}\frac{B_{2k}}{(2k)!}\left(F^{(2k-1)}(n)-F^{(2k-1)}(0)\right)
+{\rm remainder}
\ee
where $B_{i}$ are the Bernoulli numbers. 
After collecting a volume contribution, the only terms we need to compute are the $0$-th and $1$-st power of $R$.
Note that only the integral depends on $R$, and therefore, in dimensions $d>2$
for the terms that we are interested in it is enough to compute the integral.

Since the volume of the $d-$sphere is $V(d)=\frac{2}{d!} \Gamma\left(\frac{d}{2}+1\right) \left(4\pi\right)^{d/2}\left(\frac{(d-1)d}{R}\right)^{d/2}$
we only have to isolate the terms in the integral proportional to $R^{-d/2}$ and $R^{1-d/2}$
\be
2\,2^{\left[\frac{d}{2}\right]}\int_{0}^{k\sqrt{\frac{d(d-1)}{R}}-\frac{d}{2}}\mathrm{d}n\ \left(\begin{array}{c}
n+d-1\\
n
\end{array}\right)=2\,\frac{2^{\left[\frac{d}{2}\right]}}{\left(d-1\right)!}\int_{0}^{k\sqrt{\frac{d(d-1)}{R}}-\frac{d}{2}}\mathrm{d}n\ \left(n+d-1\right)\cdots\left(n+1\right)
\ee
changing variables $n\to n'-d/2$
\be
\label{longpolynomial}
2\frac{2^{\left[\frac{d}{2}\right]}}{\left(d-1\right)!}\int_{\frac{d}{2}}^{k\sqrt{\frac{d(d-1)}{R}}}\mathrm{d}n'\ \left(n'+\frac{d}{2}-1\right)\cdots\left(n'-\left(\frac{d}{2}-1\right)\right)
\ee
the terms we are interested in come from the integral of the two highest
order power of $n'$
\be
\left(n'+\frac{d}{2}-1\right)\cdots\left(n'-\left(\frac{d}{2}-1\right)\right)=n'^{d-1}-n'^{d-3}\sum_{k=1}^{\left[\frac{d-1}{2}\right]}\left(\frac{d}{2}-k\right)^{2}+\cdots
\ee
we can rewrite the sum $\sum_{k=1}^{\left[\frac{d-1}{2}\right]}\left(\frac{d}{2}-k\right)^{2}=\frac{1}{24}d\left(d-1\right)(d-2)$,
and perform the integral
\be
\mathrm{Tr}\left[\frac{\partial_{t}R_{k}}{P_{k}}\right]=2\frac{2^{\left[\frac{d}{2}\right]}}{\left(d-1\right)!}\frac{1}{d}\left(k\sqrt{\frac{d(d-1)}{R}}\right)^{d}-2\frac{2^{\left[\frac{d}{2}\right]}}{\left(d-1\right)!}\frac{1}{d-2}\left(k\sqrt{\frac{d(d-1)}{R}}\right)^{d-2}\frac{1}{24}d\left(d-1\right)(d-2)+\cdots
\ee
Collecting the volume of $\mathcal{S}^{d}$ we obtain
\be
\frac{d\Gamma_k}{dt}=
-\mathrm{Tr}\left[\frac{\partial_{t}R_{k}}{P_{k}}\right]=
-\frac{1}{\Gamma\left(\frac{d}{2}+1\right)}
\frac{2^{\left[\frac{d}{2}\right]}}{\left(4\pi\right)^{\frac{d}{2}}}
V(d)
\left(k^{d}-\frac{d}{24}k^{d-2}R+O\left(R^{2}\right)\right)
\ee

\section{Appendix II: Type Ib calculation}

We report here the detailed computation of the $A$ and $B$ coefficients of (\ref{eqconf1},\ref{eqconf2}) 
for a Type Ib cutoff. 
The FRGE is the sum of traces over the irreducible components of the metric fluctuation
defined in \eqref{TTdecomposition}. They give: 
\begin{small}
\begin{align}
&\hspace{-0.25cm} \frac{1}{2}\mathrm{Tr}_{(2)}\frac{\partial_{t}R_{k}+\eta R_{k}}{P_{k}+\left(\frac{d\left(d-3\right)+4}{d\left(d-1\right)}-\zeta\frac{d-2}{2d}\right)R-(2-\zeta)\Lambda}=
\\
&\hspace{-0.25cm}\frac{1}{2}\frac{1}{\left(4\pi\right)^{d/2}}\int dx\sqrt{g}
\Bigg[\frac{(d-2)(d+1)}{2}Q_{\frac{d}{2}}\left(\frac{\partial_{t}R_{k}+\eta R_{k}}{P_{k}-(2-\zeta)\Lambda}\right)
 +\frac{(d-5)(d+1)(d+2)}{12(d-1)}Q_{\frac{d}{2}-1}\left(\frac{\partial_{t}R_{k}+\eta R_{k}}{P_{k}-(2-\zeta)\Lambda}\right)R 
 \nonumber\\
&\hspace{2.8cm} -\left(\frac{d\left(d-3\right)+4}{d\left(d-1\right)}-\zeta\frac{d-2}{2d}\right)
\frac{(d-2)(d+1)}{2}Q_{\frac{d}{2}}\left(\frac{\partial_{t}R_{k}+\eta R_{k}}{\left(P_{k}-(2-\zeta)\Lambda\right)^{2}}\right)R\Bigg]
\nonumber
\end{align}
\begin{align}
&\hspace{-1.7cm} \frac{1}{2}\mathrm{Tr}'_{(1)}\frac{\partial_{t}R_{k}+\eta R_{k}}{P_{k}+\left(\frac{\alpha\left(d-2\right)-1}{d}-\zeta\alpha\frac{d-2}{2d}\right)R-\alpha\left(2-\zeta\right)\Lambda}=
\\
&\hspace{-1.7cm}\frac{1}{2}\frac{1}{\left(4\pi\right)^{d/2}}\int dx\sqrt{g}
\Bigg[(d-1)Q_{\frac{d}{2}}\left(\frac{\partial_{t}R_{k}+\eta R_{k}}{P_{k}-\alpha\left(2-\zeta\right)\Lambda}\right)
+\frac{(d-3)(d+2)}{6d}Q_{\frac{d}{2}-1}\left(\frac{\partial_{t}R_{k}+\eta R_{k}}{P_{k}-\alpha\left(2-\zeta\right)\Lambda}\right)R
  \nonumber\\
 &\hspace{1.5cm} -\left(\frac{\alpha\left(d-2\right)-1}{d}-\zeta\alpha\frac{d-2}{2d}\right)(d-1)Q_{\frac{d}{2}}\left(\frac{\partial_{t}R_{k}+\eta R_{k}}{\left(P_{k}-\alpha\left(2-\zeta\right)\Lambda\right)^{2}}\right)R\Bigg]
  \nonumber
\end{align}
\begin{align}
&\hspace{-3.2cm} \frac{1}{2}\mathrm{Tr''}_{(0)}\frac{\partial_{t}R_{k}+\eta R_{k}}{P_{k}+\frac{\alpha(d-2)(2-\zeta)-4}{2(d-1)-\alpha(d-2)}R
-\frac{\alpha d(2-\zeta)}{2\left(d-1\right)-\alpha\left(d-2\right)}\Lambda}=
\\
&\hspace{-3.2cm} \frac{1}{2}\frac{1}{\left(4\pi\right)^{d/2}}
\int dx\sqrt{g}\Bigg[
Q_{\frac{d}{2}}\left(\frac{\partial_{t}R_{k}+\eta R_{k}}{P_{k}-\frac{\alpha d(2-\zeta)}{2\left(d-1\right)-\alpha\left(d-2\right)}\Lambda}\right)
+\frac{1}{6}Q_{\frac{d}{2}-1}\left(\frac{\partial_{t}R_{k}+\eta R_{k}}{P_{k}-\frac{\alpha d(2-\zeta)}{2\left(d-1\right)-\alpha\left(d-2\right)}\Lambda}\right)R
\nonumber\\
&\hspace{0.cm} -\frac{\alpha(d-2)(2-\zeta)-4}{2(d-1)-\alpha(d-2)}
Q_{\frac{d}{2}}\left(\frac{\partial_{t}R_{k}+\eta R_{k}}{\left(P_{k}-\frac{\alpha d(2-\zeta)}{2\left(d-1\right)-\alpha\left(d-2\right)}\Lambda\right)^{2}}\right)R\Bigg]
\nonumber
\end{align}
\begin{align}
&\hspace{-3.2cm}\frac{1}{2}\mathrm{Tr}_{(0)}\frac{\partial_{t}R_{k}+\eta R_{k}}{P_{k}+\frac{d-4+\zeta}{2(d-1)-\alpha(d-2)}R-\frac{2d\left(1+\frac{\zeta}{d-2}\right)}{2(d-1)-\alpha(d-2)}\Lambda}=\\
&\hspace{-3.2cm}
\frac{1}{2}\frac{1}{\left(4\pi\right)^{d/2}}\int dx\sqrt{g}
\Bigg[Q_{\frac{d}{2}}\left(\frac{\partial_{t}R_{k}+\eta R_{k}}{P_{k}-\frac{2d\left(1+\frac{\zeta}{d-2}\right)}{2(d-1)-\alpha(d-2)}\Lambda}\right)
+\frac{1}{6}Q_{\frac{d}{2}-1}\left(\frac{\partial_{t}R_{k}+\eta R_{k}}{P_{k}-\frac{2d\left(1+\frac{\zeta}{d-2}\right)}{2(d-1)-\alpha(d-2)}\Lambda}\right)R
\nonumber\\
&-\frac{d-4+\zeta}{2(d-1)-\alpha(d-2)}Q_{\frac{d}{2}}\left(\frac{\partial_{t}R_{k}+\eta R_{k}}{\left(P_{k}-\frac{2d\left(1+\frac{\zeta}{d-2}\right)}{2(d-1)-\alpha(d-2)}\Lambda\right)^{2}}\right)R\Bigg]
\nonumber
\end{align}
\end{small}
Here a prime or a double prime indicate that the first or the first and the second eigenvalues have to be omitted from the trace (because $\xi_\mu$ and $\sigma$ obey to some differential constraints, for more details see for example \cite{cpr2}).
The contribution of the transverse and longitudinal parts of the diffeomorphism ghosts are 
\begin{small}
\begin{align}
-\mathrm{Tr}{}_{(1)}\frac{\partial_{t}R_{k}}{P_{k}-\frac{R}{d}}= & 
-\frac{1}{\left(4\pi\right)^{d/2}}\int dx\sqrt{g}
\Bigg[(d-1)Q_{\frac{d}{2}}\left(\frac{\partial_{t}R_{k}}{P_{k}}\right)
-\frac{d-1}{d}Q_{\frac{d}{2}}\left(\frac{\partial_{t}R_{k}}{P_{k}^{2}}\right)R
\nonumber\\
&\hspace{4.4cm}
+\frac{\left(d+2\right)\left(d-3\right)}{6d}
Q_{\frac{d}{2}-1}\left(\frac{\partial_{t}R_{k}}{P_{k}}\right)R\Bigg]\ ;\\
-\mathrm{Tr}{}_{(0)}\frac{\partial_{t}R_{k}}{P_{k}-\frac{2R}{d}}= & 
-\frac{1}{\left(4\pi\right)^{d/2}}\int dx\sqrt{g}
\Bigg[Q_{\frac{d}{2}}\left(\frac{\partial_{t}R_{k}}{P_{k}}\right)-\frac{2}{d}Q_{\frac{d}{2}}\left(\frac{\partial_{t}R_{k}}{P_{k}^{2}}\right)R
+\frac{1}{6}Q_{\frac{d}{2}-1}\left(\frac{\partial_{t}R_{k}}{P_{k}}\right)R\Bigg]\ .
\end{align}
\end{small}
Collecting the coefficients of $\int\sqrt{g}$ and $-\int\sqrt{g}R$ we 
extract the $A$ and $B$ coefficients:
\begin{small}
\begin{align}
\label{A1}
A_1= \frac{1}{2}\frac{16\pi}{\left(4\pi\right)^{d/2}}\Bigg[&\frac{(d-2)(d+1)}{2}
\tilde Q_{\frac{d}{2}}\left(\frac{\partial_{t}R_{k}}{P_{k}-(2-\zeta)\Lambda}\right)
+(d-1)\tilde Q_{\frac{d}{2}}\left(\frac{\partial_{t}R_{k}}{P_{k}-\alpha(2-\zeta)\Lambda}\right)
\nonumber\\
& \hspace{-1cm}
+\tilde Q_{\frac{d}{2}}
\left(\frac{\partial_{t}R_{k}}{P_{k}-\frac{\alpha d(2-\zeta)}{2\left(d-1\right)-\alpha\left(d-2\right)}\Lambda}\right)
+\tilde Q_{\frac{d}{2}}\left(
\frac{\partial_{t}R_{k}}{P_{k}-\frac{2d\left(1+\frac{\zeta}{d-2}\right)}{2(d-1)-\alpha(d-2)}\Lambda}\right)
-2d\tilde Q_{\frac{d}{2}}\left(\frac{\partial_{t}R_{k}}{P_{k}}\right)
\Bigg]
\\
\label{A2}
A_2=  \frac{1}{2}\frac{16\pi}{\left(4\pi\right)^{d/2}}\Bigg[&
\frac{(d-2)(d+1)}{2}\tilde Q_{\frac{d}{2}}\left(\frac{R_{k}}{P_{k}-(2-\zeta)\Lambda}\right)
+(d-1)\tilde Q_{\frac{d}{2}}\left(\frac{R_{k}}{P_{k}-\alpha(2-\zeta)\Lambda}\right)
\nonumber\\
&  \hspace{-1cm}
+\tilde Q_{\frac{d}{2}}
\left(\frac{R_{k}}{P_{k}-\frac{\alpha d(2-\zeta)}{2\left(d-1\right)-\alpha\left(d-2\right)}\Lambda}\right)
+\tilde Q_{\frac{d}{2}}\left(
\frac{R_{k}}{P_{k}-\frac{2d\left(1+\frac{\zeta}{d-2}\right)}{2(d-1)-\alpha(d-2)}\Lambda}\right)
\Bigg]
\\
B_1=  \frac{1}{2}\frac{16\pi}{\left(4\pi\right)^{d/2}}\Bigg[&
\frac{(d-5)(d+1)(d+2)}{12(d-1)}\tilde Q_{\frac{d}{2}-1}\left(\frac{\partial_{t}R_{k}}{P_{k}-(2-\zeta)\Lambda}\right)
\nonumber\\
& \hspace{-1cm}
-\left(\frac{d\left(d-3\right)+4}{d\left(d-1\right)}-\zeta\frac{d-2}{2d}\right)\frac{(d-2)(d+1)}{2}
\tilde Q_{\frac{d}{2}}\left(\frac{\partial_{t}R_{k}}{\left(P_{k}-(2-\zeta)\Lambda\right)^{2}}\right)
\nonumber\\
& \hspace{-1cm}
+\frac{(d-3)(d+2)}{6d}
\tilde Q_{\frac{d}{2}-1}\left(\frac{\partial_{t}R_{k}}{P_{k}-\alpha(2-\zeta)\Lambda}\right)
\nonumber\\
& \hspace{-1cm}
-\left(\frac{\alpha\left(d-2\right)-1}{d}-\zeta\alpha\frac{d-2}{2d}\right)(d-1)
\tilde Q_{\frac{d}{2}}\left(\frac{\partial_{t}R_{k}}{\left(P_{k}-\alpha(2-\zeta)\Lambda\right)^{2}}\right)
\nonumber\\
& \hspace{-1cm}
+\frac{1}{6}\tilde Q_{\frac{d}{2}-1}\left(\frac{\partial_{t}R_{k}}{P_{k}-\frac{\alpha d(2-\zeta)}{2\left(d-1\right)
-\alpha\left(d-2\right)}\Lambda}\right)
-\frac{\alpha(d-2)(2-\zeta)-4}{2(d-1)-\alpha(d-2)}
\tilde Q_{\frac{d}{2}}\left(\frac{\partial_{t}R_{k}}{\left(P_{k}-\frac{\alpha d(2-\zeta)}{2\left(d-1\right)-\alpha\left(d-2\right)}\Lambda\right)^{2}}\right)
\nonumber\\
 &\hspace{-1cm}
+\frac{1}{6}
\tilde Q_{\frac{d}{2}-1}\left(\frac{\partial_{t}R_{k}}{P_{k}-\frac{2d\left(1+\frac{\zeta}{d-2}\right)}{2(d-1)-\alpha(d-2)}\Lambda}\right)
-\frac{d-4+\zeta}{2(d-1)-\alpha(d-2)}
\tilde Q_{\frac{d}{2}}\left(\frac{\partial_{t}R_{k}}{\left(P_{k}-\frac{2d\left(1+\frac{\zeta}{d-2}\right)}{2(d-1)-\alpha(d-2)}\Lambda\right)^{2}}\right)
\nonumber \\
 &\hspace{-1cm} -\frac{d^{2}-6}{3d}
 \tilde Q_{\frac{d}{2}-1}\left(\frac{\partial_{t}R_{k}}{P_{k}}\right)
 -2\frac{d+1}{d}\tilde Q_{\frac{d}{2}}\left(\frac{\partial_{t}R_{k}}{P_{k}^{2}}\right) 
\Bigg]
\end{align}
\begin{align}
B_2=  \frac{1}{2}\frac{16\pi}{\left(4\pi\right)^{d/2}}\Bigg[&
\frac{(d-5)(d+1)(d+2)}{12(d-1)}
\tilde Q_{\frac{d}{2}-1}\left(\frac{R_{k}}{P_{k}-(2-\zeta)\Lambda}\right)
\nonumber\\
&\hspace{-1cm}
-\left(\frac{d\left(d-3\right)+4}{d\left(d-1\right)}-\zeta\frac{d-2}{2d}\right)\frac{(d-2)(d+1)}{2}
\tilde Q_{\frac{d}{2}}\left(\frac{R_{k}}{\left(P_{k}-(2-\zeta)\Lambda\right)^{2}}\right)
\nonumber\\
 & \hspace{-1cm}
 +\frac{(d-3)(d+2)}{6d}\tilde Q_{\frac{d}{2}-1}\left(\frac{R_{k}}{P_{k}-\alpha(2-\zeta)\Lambda}\right)
\nonumber\\
&\hspace{-1cm}-\left(\frac{\alpha\left(d-2\right)-1}{d}-\zeta\alpha\frac{d-2}{2d}\right)(d-1)
\tilde Q_{\frac{d}{2}}\left(\frac{R_{k}}{\left(P_{k}-\alpha(2-\zeta)\Lambda\right)^{2}}\right)
\nonumber\\
& \hspace{-1cm}+\frac{1}{6}
\tilde Q_{\frac{d}{2}-1}\left(\frac{R_{k}}{P_{k}-\frac{\alpha d(2-\zeta)}{2\left(d-1\right)
-\alpha\left(d-2\right)}\Lambda}\right)
-\frac{\alpha(d-2)(2-\zeta)-4}{2(d-1)-\alpha(d-2)}
\tilde Q_{\frac{d}{2}}\left(\frac{R_{k}}{\left(P_{k}-\frac{\alpha d(2-\zeta)}{2\left(d-1\right)-\alpha\left(d-2\right)}\Lambda\right)^{2}}\right)
\nonumber\\
 & \hspace{-1cm}
+\frac{1}{6}
\tilde Q_{\frac{d}{2}-1}\left(\frac{R_{k}}{P_{k}-\frac{2d\left(1+\frac{\zeta}{d-2}\right)}{2(d-1)-\alpha(d-2)}\Lambda}\right)
-\frac{d-4+\zeta}{2(d-1)-\alpha(d-2)}
\tilde Q_{\frac{d}{2}}\left(\frac{R_{k}}{\left(P_{k}-\frac{2d\left(1+\frac{\zeta}{d-2}\right)}{2(d-1)-\alpha(d-2)}\Lambda\right)^{2}}\right)
\Bigg]
\end{align}
\end{small}
Here we have defined the dimensionless versions of the $Q$-functionals:
$\tilde Q_\frac{d}{2}=k^{-d}Q_\frac{d}{2}$ and
$\tilde Q_{\frac{d}{2}-1}=k^{2-d}Q_{\frac{d}{2}-1}$.

Finally let us consider the contribution of Lorentz ghosts.
They do not propagate and therefore are usually neglected in the evaluation of the effective action
in perturbation theory.
Nevertheless if, following \cite{hr}, we impose a cutoff on their determinant
they contribute to the r.h.s. of the FRGE an amount
\be
\label{LGhost}
-\mathrm{Tr}\frac{\partial_{t}R_{k}}{R_{k}+\frac{2\mu^{2}}{\sqrt\zeta}}
=-\frac{1}{\left(4\pi\right)^{d/2}}\frac{d(d-1)}{2}\int dx\sqrt{g}\
\left[Q_{\frac{d}{2}}\left(\frac{\partial_{t}R_{k}}{R_{k}+\frac{2\mu^{2}}{\sqrt\zeta}}\right)
+\frac{1}{6}Q_{\frac{d}{2}-1}\left(\frac{\partial_{t}R_{k}}{R_{k}+\frac{2\mu^{2}}{\sqrt\zeta}}\right)R\right]
\ee
Here we have introduced the arbitrary mass parameter $\mu$ (denoted $\bar\mu$ in \cite{hr}).
In particular note that in the limit $\mu\to\infty$ the ghost contribution vanishes
and one recovers the standard perturbative result where the Lorentz ghosts are neglected.
Let $A^L$ and $B^L$ be the contribution of the Lorentz ghosts to the coefficients $A_1$ and $B_1$,
defined in (\ref{eqconf1},\ref{eqconf2}).
From the above we read off
\be
A^L=-\frac{16\pi}{\left(4\pi\right)^{d/2}}\frac{d(d-1)}{2}
\tilde Q_{\frac{d}{2}}\left(\frac{\partial_{t}R_{k}}{R_{k}+\frac{2\mu^{2}}{\sqrt\zeta}}\right)\ ;
\qquad
B^L=-\frac{16\pi}{\left(4\pi\right)^{d/2}}\frac{d(d-1)}{12}
\tilde Q_{\frac{d}{2}-1}\left(\frac{\partial_{t}R_{k}}{R_{k}+\frac{2\mu^{2}}{\sqrt\zeta}}\right)\ .
\ee
Note the appearance of $R_k$ instead of $P_k$ in the denominators.
In general the $Q$ functionals $Q_{n}\left(\frac{\partial_{t}R_{k}}{R_{k}+2\mu^{2}/\sqrt{\zeta}}\right)$
can be computed explicitly, with cutoff (\ref{opt}), in terms of hypergeometric functions.
For the calculations in four dimensions we only need the following
\begin{small}
\begin{eqnarray}
&&
\tilde Q_{1}\left(\frac{\partial_{t}R_{k}}{R_{k}+\frac{2\mu^{2}}{\sqrt{\zeta}}}\right)=
{\rm Log}\left(1+\frac{\sqrt{\zeta}}{2\tilde\mu^{2}}\right)\ ;
\\
&&
\tilde Q_{2}\left(\frac{\partial_{t}R_{k}}{R_{k}+\frac{2\mu^{2}}{\sqrt{\zeta}}}\right)=
-1+\left(1+\frac{2\tilde\mu^{2}}{\sqrt{\zeta}}\right)
{\rm Log}\left(1+\frac{\sqrt{\zeta}}{2\tilde\mu^{2}}\right)\ .
\end{eqnarray}
\end{small}
\section{Appendix III: Type IIb calculation}
We report here the $A$ and $B$ coefficients of (\ref{eqconf1},\ref{eqconf2}) 
for a Type IIb cutoff. 
The contributions of the irreducible components of the metric fluctuation
to the r.h.s. of the FRGE are
\begin{small}
\begin{align}
\hspace{-2cm}\frac{1}{2}\mathrm{Tr}_{(2)}\frac{\partial_{t}R_{k}+\eta R_{k}}{P_{k}-(2-\zeta)\Lambda}=&  \frac{1}{2}\frac{1}{\left(4\pi\right)^{d/2}}\int dx\sqrt{g}
\Bigg[\frac{(d-2)(d+1)}{2}Q_{\frac{d}{2}}\left(\frac{\partial_{t}R_{k}+\eta R_{k}}{P_{k}-(2-\zeta)\Lambda}\right)\\
 &
 -\frac{(d+1)}{12d}\left(5d^{2}-22d+48-3\zeta\left(d-2\right){}^{2}\right)\ Q_{\frac{d}{2}-1}\left(\frac{\partial_{t}R_{k}+\eta R_{k}}{P_{k}-(2-\zeta)\Lambda}\right)R\Bigg]
 \nonumber
\end{align}
\begin{align}
\hspace{-0.9cm}\frac{1}{2}\mathrm{Tr}'_{(1)}\frac{\partial_{t}R_{k}+\eta R_{k}}{P_{k}-\alpha\left(2-\zeta\right)\Lambda}=& \frac{1}{2}\frac{1}{\left(4\pi\right)^{d/2}}\int dx\sqrt{g}
\Bigg[
(d-1)Q_{\frac{d}{2}}\left(\frac{\partial_{t}R_{k}+\eta R_{k}}{P_{k}-\alpha\left(2-\zeta\right)\Lambda}\right)
\\
&- \frac{\left(d^{2}+5d-12+3\alpha(\zeta-2)\left(d-1\right)\left(d-2\right)\right) }{6d}\ Q_{\frac{d}{2}-1}\left(\frac{\partial_{t}R_{k}+\eta R_{k}}{P_{k}-\alpha\left(2-\zeta\right)\Lambda}\right)R\Bigg]
\nonumber
\end{align}
\begin{align}
\hspace{-1.6cm}\frac{1}{2}\mathrm{Tr''}_{(0)}\frac{\partial_{t}R_{k}+\eta R_{k}}{P_{k}-\frac{\alpha d(2-\zeta)}{2\left(d-1\right)-\alpha\left(d-2\right)}\Lambda}= & 
\frac{1}{2}\frac{1}{\left(4\pi\right)^{d/2}}\int dx\sqrt{g}
\Bigg[Q_{\frac{d}{2}}\left(\frac{\partial_{t}R_{k}+\eta R_{k}}{P_{k}-\frac{\alpha d(2-\zeta)}{2\left(d-1\right)-\alpha\left(d-2\right)}\Lambda}\right)
\\
&+\frac{14+(2-7\alpha-3\zeta\alpha)(d-2)}{6(2(d-1)-\alpha(d-2))}
Q_{\frac{d}{2}-1}\left(\frac{\partial_{t}R_{k}+\eta R_{k}}{P_{k}-\frac{\alpha d(2-\zeta)}{2\left(d-1\right)-\alpha\left(d-2\right)}\Lambda}\right)R \Bigg]
\nonumber
\end{align}
\begin{align}
\hspace{-2.2cm}\frac{1}{2}\mathrm{Tr}_{(0)}\frac{\partial_{t}R_{k}+\eta R_{k}}{P_{k}-\frac{2d\left(1+\frac{\zeta}{d-2}\right)}{2(d-1)-\alpha(d-2)}\Lambda}= &\frac{1}{2}\frac{1}{\left(4\pi\right)^{d/2}}\int dx\sqrt{g}
\Bigg[Q_{\frac{d}{2}}\left(\frac{\partial_{t}R_{k}+\eta R_{k}}{P_{k}-\frac{2d\left(1+\frac{\zeta}{d-2}\right)}{2(d-1)-\alpha(d-2)}\Lambda}\right)
\\
&
+\frac{22-4d+(2-d)\alpha-6\zeta}{6(2(d-1)-\alpha(d-2))}
Q_{\frac{d}{2}-1}\left(\frac{\partial_{t}R_{k}+\eta R_{k}}{P_{k}
-\frac{2d\left(1+\frac{\zeta}{d-2}\right)}{2(d-1)-\alpha(d-2)}\Lambda}\right) R\Bigg]
\nonumber
\end{align}
\end{small}
The contribution of the diffeomorphism ghosts is
\begin{small}
\begin{align}
-\mathrm{Tr}_{(1)}\frac{\partial_{t}R_{k}}{P_{k}}= & 
-\frac{1}{\left(4\pi\right)^{d/2}}\int dx\sqrt{g}
\left[(d-1)Q_{\frac{d}{2}}\left(\frac{\partial_{t}R_{k}}{P_{k}}\right)
+\frac{d^2+5d-12}{6d}R\,Q_{\frac{d}{2}-1}\left(\frac{\partial_{t}R_{k}}{P_{k}}\right)\right]
\\
-\mathrm{Tr}_{(0)}\frac{\partial_{t}R_{k}}{P_{k}}= & 
-\frac{1}{\left(4\pi\right)^{d/2}}\int dx\sqrt{g}
\left[Q_{\frac{d}{2}}\left(\frac{\partial_{t}R_{k}}{P_{k}}\right)
+\frac{d+12}{6d}R\,Q_{\frac{d}{2}-1}\left(\frac{\partial_{t}R_{k}}{P_{k}}\right)\right]\ .
\end{align}
\end{small}
The contribution of Lorentz ghosts is the same as in the Type Ib case.
The coefficients $A_1$ and $A_2$ are the same as in equations (\ref{A1}, \ref{A2}), whereas
\begin{small}
\begin{align}
\hspace{-0.3cm}B_1=  \frac{1}{2}\frac{16\pi}{\left(4\pi\right)^{d/2}}
\Bigg[&-\frac{(d+1)}{12d}\left(5d^{2}-22d+48-3\zeta\left(d-2\right){}^{2}\right) 
\tilde Q_{\frac{d}{2}-1}\left(\frac{\partial_{t}R_{k}}{P_{k}-(2-\zeta)\Lambda}\right)
\\
& 
+\frac{d^2+5d-12+3\alpha(\zeta-2)(d-1)(d-2)}{6d} 
\tilde Q_{\frac{d}{2}-1}\left(\frac{\partial_{t}R_{k}}{P_{k}-\alpha\left(2-\zeta\right)\Lambda}\right)
\nonumber\\
 & 
+\frac{22-4d+(2-d)\alpha-6\zeta}{6(2(d-1)-\alpha(d-2))}
\tilde Q_{\frac{d}{2}-1}\left(\frac{\partial_{t}R_{k}}{P_{k}-\frac{2d\left(1+\frac{\zeta}{d-2}\right)}{2(d-1)-\alpha(d-2)}\Lambda}\right)
\nonumber\\
& 
+\frac{14+(2-7\alpha-3\zeta\alpha)(d-2)}{6(2(d-1)-\alpha(d-2))} 
\tilde Q_{\frac{d}{2}-1}\left(\frac{\partial_{t}R_{k}}{P_{k}-\frac{\alpha d(2-\zeta)}{2\left(d-1\right)-\alpha\left(d-2\right)}\Lambda}\right)
-\frac{d+6}{3}\ \tilde Q_{\frac{d} {2}-1}\left(\frac{\partial_{t}R_{k}}{P_{k}}\right)
\Bigg]
\nonumber\\
B_2=\frac{1}{2}\frac{16\pi}{\left(4\pi\right)^{d/2}}
\Bigg[&
-\frac{(d+1)}{12d}\left(5d^{2}-22d+48-3\zeta\left(d-2\right){}^{2}\right) 
\tilde Q_{\frac{d}{2}-1}\left(\frac{R_{k}}{P_{k}-(2-\zeta)\Lambda}\right)
\nonumber\\
& 
+\frac{d^2+5d-12+3\alpha(\zeta-2)(d-1)(d-2)}{6d} 
\tilde Q_{\frac{d}{2}-1}\left(\frac{R_{k}}{P_{k}-\alpha\left(2-\zeta\right)\Lambda}\right)
\nonumber\\
 & 
+\frac{22-4d+(2-d)\alpha-6\zeta}{6(2(d-1)-\alpha(d-2))} 
\tilde Q_{\frac{d}{2}-1}\left(\frac{R_{k}}{P_{k}-\frac{2d\left(1+\frac{\zeta}{d-2}\right)}{2(d-1)-\alpha(d-2)}\Lambda}\right)
\nonumber\\
& 
+\frac{14+(2-7\alpha-3\zeta\alpha)(d-2)}{6(2(d-1)-\alpha(d-2))}\ 
\tilde Q_{\frac{d}{2}-1}\left(\frac{R_{k}}{P_{k}-\frac{\alpha d(2-\zeta)}{2\left(d-1\right)-\alpha\left(d-2\right)}\Lambda}\right)
\Bigg]
\end{align}
\end{small}


\end{document}